\PassOptionsToPackage{unicode}{hyperref}
\PassOptionsToPackage{hyphens}{url}
\documentclass[
]{article}
\usepackage{lmodern}
\usepackage{amsmath}
\usepackage{ifxetex,ifluatex}
\ifnum 0\ifxetex 1\fi\ifluatex 1\fi=0 
  \usepackage[T1]{fontenc}
  \usepackage[utf8]{inputenc}
  \usepackage{textcomp} 
  \usepackage{amssymb}
\else 
  \usepackage{unicode-math}
  \defaultfontfeatures{Scale=MatchLowercase}
  \defaultfontfeatures[\rmfamily]{Ligatures=TeX,Scale=1}
\fi
\IfFileExists{upquote.sty}{\usepackage{upquote}}{}
\IfFileExists{microtype.sty}{
  \usepackage[]{microtype}
  \UseMicrotypeSet[protrusion]{basicmath} 
}{}
\makeatletter
\@ifundefined{KOMAClassName}{
  \IfFileExists{parskip.sty}{%
    \usepackage{parskip}
  }{
    \setlength{\parindent}{0pt}
    \setlength{\parskip}{6pt plus 2pt minus 1pt}}
}{
  \KOMAoptions{parskip=half}}
\makeatother
\usepackage{xcolor}
\IfFileExists{xurl.sty}{\usepackage{xurl}}{} 
\IfFileExists{bookmark.sty}{\usepackage{bookmark}}{\usepackage{hyperref}}
\hypersetup{
  pdftitle={Ethical Assurance},
  pdfauthor={Christopher Burr; David Leslie},
  pdfkeywords={argument-based assurance, machine learning, responsible
research and innovation, data ethics, argumentation theory},
  hidelinks,
  pdfcreator={LaTeX via pandoc}}
\usepackage{graphicx}
\makeatletter
\def\maxwidth{\ifdim\Gin@nat@width>\linewidth\linewidth\else\Gin@nat@width\fi}
\def\maxheight{\ifdim\Gin@nat@height>\textheight\textheight\else\Gin@nat@height\fi}
\makeatother
\setkeys{Gin}{width=\maxwidth,height=\maxheight,keepaspectratio}
\makeatletter
\def\fps@figure{htbp}
\makeatother
\providecommand{\tightlist}{%
  \setlength{\itemsep}{0pt}\setlength{\parskip}{0pt}}
\ifluatex
  \usepackage{selnolig}  
\fi
\newlength{\cslhangindent}
\newlength{\csllabelwidth}
\newenvironment{CSLReferences}[2] 
 {
  \setlength{\parindent}{0pt}
  \ifodd #1 \everypar{\setlength{\hangindent}{\cslhangindent}}\ignorespaces\fi
  \ifnum #2 > 0
  \setlength{\parskip}{#2\baselineskip}
  \fi
 }%
 {}
\usepackage{calc}

\title{Ethical Assurance}
\usepackage{etoolbox}
\makeatletter
\providecommand{\subtitle}[1]{
  \apptocmd{\@title}{\par {\large #1 \par}}{}{}
}
\makeatother
\subtitle{A practical approach to the responsible design, development,
and deployment of data-driven technologies}
\author{Christopher Burr\footnote{The Alan Turing Institute, UK
  \textbar{} Corresponding author: cburr@turing.ac.uk} \and David
Leslie\footnote{The Alan Turing Institute, UK}}
\date{\today}

\begin{document}
\maketitle
\begin{abstract}
This article offers several contributions to the interdisciplinary
project of responsible research and innovation in data science and AI.
First, it provides a critical analysis of current efforts to establish
practical mechanisms for algorithmic assessment, which are used to
operationalise normative principles, such as sustainability,
accountability, transparency, fairness, and explainability, in order to
identify limitations and gaps with the current approaches. Second, it
provides an accessible introduction to the methodology of argument-based
assurance, and explores how it is currently being applied in the
development of safety cases for autonomous and intelligent systems.
Third, it generalises this method to incorporate wider ethical, social,
and legal considerations, in turn establishing a novel version of
argument-based assurance that we call `ethical assurance.' Ethical
assurance is presented as a structured means for unifying the myriad
practical mechanisms that have been proposed, as it is built upon a
process-based form of project governance that supports inclusive and
participatory ethical deliberation while also remaining grounded in
social and technical realities. Finally, it sets an agenda for ethical
assurance, by detailing current challenges, open questions, and next
steps, which serve as a springboard to build an active (and
interdisciplinary) research programme as well as contribute to ongoing
discussions in policy and governance.
\end{abstract}

\hypertarget{introduction}{%
\section{Introduction}\label{introduction}}

\hypertarget{setting-the-stage}{%
\subsection{Setting the Stage}\label{setting-the-stage}}

The recent history of artificial intelligence (AI) ethics and governance
has been characterised by increasingly vocal calls for a move from
\emph{principles to practice}. Over the past several years, some have
discerned a rapid transition in the field from an initial concentration
on high-level principles and techno-solutionist ``fixes'' (e.g.~for
issues such as algorithmic bias) towards a ``third wave'' of hard-nosed
advocacy and legal action that is focused on ``practical mechanisms for
rectifying power imbalances and achieving individual and societal
justice'' (\protect\hyperlink{ref-kind2020}{Kind 2020}). Others have
emphasised that ``closing the gap'' between principles and practice
should involve the employment of myriad tools and methods throughout the
various stages of the a project's lifecycle, so that the ``what'' of
ethical principles can be translated into the ``how'' of ``technical
mechanisms'' (\protect\hyperlink{ref-morley2019f}{Morley et al. 2019}).
Others still have called for a strengthening of regimes of
``auditability,'' ``traceability,'' and ``reviewability,'' emphasising
the importance of oversight, accountability, and transparency as the key
to the effective governance for responsible AI research and innovation
(\protect\hyperlink{ref-cobbe2021}{Cobbe, Lee, and Singh 2021};
\protect\hyperlink{ref-kroll2021}{Kroll 2021};
\protect\hyperlink{ref-mokander2021}{Mökander and Floridi 2021};
\protect\hyperlink{ref-raji2020}{Raji et al. 2020}).

Notwithstanding the substantial merits of this intensifying
concentration on the nexus between moral concepts and social praxis,
each of these perspectives has fallen short of fully realising the
transformations they identify and enjoin. For instance, while Kind's
(\protect\hyperlink{ref-kind2020}{2020}) three waves historiography of
AI ethics has intuitive curb appeal, it may well run the risk of
reflecting feedback loops of Twitter trends and conference culture
rather than explicating the longer-term development of applied ethics in
the domain of data-driven innovation. Rather than originating solely
from the ethereal thinking of philosophers over the past few years, the
evolution of the latter has, in fact, a much deeper history. It is a
history that is deeply interwoven with emerging traditions of bioethics
and human rights regimes that arose as direct responses to tangible,
technologically inflicted harms and atrocities and in contexts of the
struggles against technological power (Leslie, 2020). In a significant
sense, that is, both traditions stem from concerted public acts of
resistance against violence done to disempowered or vulnerable
people---violence typified by the well-known barbarisms and genocides of
the mid-twentieth century, in the case of human rights, and several
atrocities of human experimentation (such as the infamous Tuskegee
syphilis experiment) that spurred the development of bioethics.

Furthermore, the emergence of critical and legal perspectives on
rectifying power imbalances and social inequities in the design and use
of algorithmic or machine learning (ML) systems has not really
represented a radical departure from the abstract musings of
philosophers and ethicists concerned with engineering robust normative
principles. Rather, it has largely been built on the normative
foundations that they have long laboured to establish. Indeed, the
demand for justice and the force of law, as these apply to the
regulation and governance of AI technologies, rely on normative
structures as their \emph{conditions of legitimacy} with values and
principles like dignity, respect, fairness, and equity providing
justificatory support, which constitutes the connective tissue that
links legal facts and juridically actionable norms
(\protect\hyperlink{ref-habermas1998}{Habermas 1998}). Therefore,
instead of viewing AI ethics as a sort of teleological movement from
superficial philosophies to the real, hands-on business of ``just AI''
(\protect\hyperlink{ref-kind2020}{Kind 2020}), it may be more accurate
and informative to see the critical, normative, technical, and juridical
aspects of AI ethics and governance as interlocking dimensions of a
longer-term, more holistic, and at times more fragile societal endeavour
to reflexively steer technology in accordance with shared human values
and purposes.

Those who have turned to the incorporation of a patchwork of technical
tools and documentation methods into the various stages of the AI/ML
project lifecycle have faced similar shortcomings. The problem here is
not that tools and method like Datasheets
(\protect\hyperlink{ref-gebru2019}{Gebru et al. 2019}), Data Nutrition
Labels (\protect\hyperlink{ref-holland2018}{Holland et al. 2018}) Data
Statements (\protect\hyperlink{ref-bender2018}{Bender and Friedman
2018}), Model Cards (\protect\hyperlink{ref-mitchell2019}{Mitchell et
al. 2019}), and FactSheets (\protect\hyperlink{ref-arnold2019}{Arnold et
al. 2019}) are of \emph{no use} as provisional attempts at closing the
gap between principles and practice, but rather that they are
\emph{nowhere near enough}. Beyond off-the-shelf tools and governance
instruments, closing the gap between principles and practice requires a
transformation of organisational cultures, technical approaches, and
individual attitudes from inside the processes and practices of design,
development, and deployment themselves. Achieving this requires
researchers, technologists, and innovators to establish and maintain
end-to-end habits of \emph{critical reflection and deliberation} across
every stage of a research or innovation project's lifecycle.

This more basic organisational, technical, and attitudinal
transformation entails that designers and developers of data-driven
technologies pay deliberate and continuous attention to the role that
values play in both discovery and engineering processes as well as in
considerations of the real-world effects that these processes yield. It
requires sustained interdisciplinary efforts to consider the
multi-faceted contexts of research and innovation, to anticipate
potential impacts, and to engage affected stakeholders inclusively in
order to ensure appropriate forms of social license and democratic
governance. An approach to building trustworthy AI/ML systems that takes
as its starting point a focus on technologically based tools or
documentation protocols (like those mentioned above) erroneously works
from the outside in, all while the actual change required to bridge the
divide between principles and practice must instead originate from
within actual research and innovation activities as part of a deeper
transformation of the organisational environments and individual
attitudes, standpoints, and dispositions whence those activities derive.

Cobbling together ``a robust `toolbox' of mechanisms to support the
verification of claims about AI systems and development processes''
(\protect\hyperlink{ref-brundage2020}{Brundage et al. 2020}), in this
latter sense, leads, in AI ethics and governance, to a kind of
functional tardiness of the governance strategies that result. Namely,
it leads to an emphasis on narrowly-targeted methods such as ``effective
assessment'' (\protect\hyperlink{ref-brundage2020}{Brundage et al.
2020}), ``auditability'' (\protect\hyperlink{ref-mokander2021}{Mökander
and Floridi 2021}; \protect\hyperlink{ref-raji2020}{Raji et al. 2020}),
``traceability'' (\protect\hyperlink{ref-kroll2021}{Kroll 2021}), and
``reviewability'' (\protect\hyperlink{ref-cobbe2021}{Cobbe, Lee, and
Singh 2021}), that show up on the scene a moment too late. Such methods
remain \emph{ex post facto} and external to the innerworkings of
sufficiently reflective and responsible modes of technology production
and use.\footnote{Test-driven development in software engineering---a
  process that uses software requirements as a basis for generating
  `test cases' prior to software development, in order to monitor and
  track the progress of the software's development---is one method that
  attempts to address this issue through the establishment of an
  internal, situated process of evaluation. Thanks to Ibrahim Habli for
  bringing this to our attention.} It is at this latter, more
foundational level of cultural formation, value shaping, and action
orientation that a bridging of the gulf between principles and practice
in AI ethics and governance must begin. In this article we propose a
first step by introducing a version of argument-based assurance that we
call `ethical assurance.'

\hypertarget{argument-based-ethical-assurance}{%
\subsection{Argument-Based (Ethical)
Assurance}\label{argument-based-ethical-assurance}}

For the purpose of this article, we offer the following definition of
argument-based assurance (ABA):

\begin{quote}
Argument-based assurance is a process of using \emph{structured
argumentation} to provide \emph{assurance} to another party (or parties)
that a particular claim (or set of related claims) about a property of a
system is warranted given the available evidence.
\end{quote}

ABA is already widely used in safety-critical domains or industries
where manufacturing and development processes are required to comply
with strict regulatory standards and support industry-recognised best
practices (\protect\hyperlink{ref-hawkins2021}{R. Hawkins et al. 2021}).
The output of this process is typically an \emph{assurance case}, which
can offer a formal, textual, or visual representation of the argument
that seeks to demonstrate how regulatory goals or standards have been
met. As we will see in \protect\hyperlink{argument-based-assurance}{§2},
this process also supports related goals such as facilitating
transparent communication and establishing trust between system or
product developers and stakeholders.

In this paper we seek to generalise the method of ABA to account for
wider normative goals, related to ethical principles such as
sustainability,accountability, fairness, or explainability. This
generalised version, known as `ethical assurance' provides a structured
method for reflecting upon how and whether normative goals have been
sufficiently established throughout the design, development, and
deployment of an AI or data-driven technology, while also facilitating a
process of active enquiry that supports meaningful stakeholder
participation and deliberation. The participatory component is necessary
for ensuring that the ethical assurance cases have moral legitimacy as
well as social license, and also helps to overcome concerns about
so-called ``ethics washing'' (\protect\hyperlink{ref-hao2019}{Hao
2019}).

In generalising ABA to accommodate wider normative considerations, we
offer a framework that is able to support anticipatory and reflective
assessment of a project's social commitments and responsibilities,
outline a procedural method for operationalising ethical principles that
results in justifiable forms of action-guidance, and address the
practical needs of technical project governance for complex data-driven
technologies.

\hypertarget{article-outlinelength}{%
\subsection[Article Outline]{\texorpdfstring{Article
Outline\footnote{This is a long article, but the sections form two
  (approximately equal) parts that focus respectively on a) laying the
  foundations for our proposal, and b) developing our proposal. The
  first part comprises §2 and §3, and the second part comprises §4 and
  §5. Furthermore, the above sections are ordered such that the earlier
  sections provide foundations for the later sections. Therefore, some
  readers may wish to skip the material that they are already familiar.
  For instance, the reader who is familiar with traditional methods of
  argument-based assurance may wish to skip
  \protect\hyperlink{argument-based-assurance}{§2} and start with the
  discussion of ML assurance in
  \protect\hyperlink{from-safety-assurance-to-ethical-assurance}{§3}.
  Our positive proposal can be found in
  \protect\hyperlink{ethical-assurance}{§4}.}}{Article Outline}}\label{article-outlinelength}}

In \protect\hyperlink{argument-based-assurance}{§2}, we introduce
argument-based assurance. Although,
\protect\hyperlink{argument-based-assurance}{§2} provides a general
introduction to argument-based assurance, it also serves as the
foundation for our own approach that is introduced starting in
\protect\hyperlink{ethical-assurance}{§4}.

In \protect\hyperlink{from-safety-assurance-to-ethical-assurance}{§3},
we address some of the limitations of existing research on ML assurance,
explain \emph{why it matters} that we address these gaps, and in turn
motivate the need for our positive proposal of ethical assurance. In
doing so, we also discuss a particular thorny issue that is regularly
observed in discussions and debates, as noted at the start of this
paper: how to operationalise ethical principles such that they provide
action-guiding constraints on practical decision-making.

With the foundations laid, in \protect\hyperlink{ethical-assurance}{§4}
we explore how to move from the methodology of \emph{safety assurance}
to \emph{ethical assurance}---a type of ABA that has been formulated
with the responsible, end-to-end governance of data-driven technologies
in mind. We present a model of a typical project lifecycle for an ML/AI
system, which is employed as a scaffold for thinking through an
illustrative example that sets out some of the activities and processes
that are required to build an \emph{ethical assurance case}. We also
explore several related topics, including how to evaluate evidence that
is employed to justify normative, goal-directed claims, and why it is
important to adopt a reflective and phased approach to the development
of ethical assurance.

In
\protect\hyperlink{conclusion-challenges-open-questions-and-next-steps}{§5},
we conclude by anticipating and responding to several challenges that
could be raised against ethical assurance, and also identify several
open issues and possible next steps for the project.

\hypertarget{argument-based-assurance}{%
\section{Argument-Based Assurance}\label{argument-based-assurance}}

In this section, we introduce \emph{argument-based assurance} (ABA) for
those readers who are unfamiliar with the methodology. However, as ABA
is the foundation upon which we build our own method, however, this
section also serves as an introduction to \emph{ethical assurance} by
introducing the reader to some of the significant functions and benefits
of ABA.

\hypertarget{what-is-argument-based-assurance}{%
\subsection{What is argument-based
assurance?}\label{what-is-argument-based-assurance}}

As already noted, ABA is a process of using \emph{structured
argumentation} to provide \emph{assurance} to another party (or parties)
that a particular claim (or set of related claims) about a property (or
properties) of a system is warranted given the available evidence. The
method is widely adopted in safety-critical domains or industries
(e.g.~automotive, energy) where manufacturing and development processes
are required to comply with strict regulatory standards and, ideally,
reflect industry-recognised best practices. A common way to meet these
compliance requirements is through the production of \emph{assurance
cases}, which provide a systematic method for justifying technical
claims regarding specific properties of a system. An assurance case can
be defined as follows:

\begin{quote}
``A reasoned and compelling argument, supported by a body of evidence,
that a system, service or organisation will operate as intended for a
defined application in a defined environment''
(\protect\hyperlink{ref-gsncommunity2018}{Community 2018, 10}).
\end{quote}

Assurance cases tend to have a particular focus or goal. For example,
the \emph{safety case} in Figure 1, which is based on an approach to
argument-based assurance known as goal-structured notation, is directed
towards and structured around a clearly defined goal (G1, top of figure)
of demonstrating that a control system is ``acceptably safe to
operate,'' within a given operating role and context (C1) (e.g.~a
component in an aircraft that will be used in well-defined environments)
(\protect\hyperlink{ref-bloomfield2010}{Bloomfield and Bishop 2010}).
Other assurance cases may focus on the security, availability, or
maintenance of a system. It is common for assurance cases to be
intricate and complex, due in part to the different relations that exist
between the various elements---depicted in Figure 1 by the different
types of arrows. However, the goal structured approach provides a clear
anchoring point

\begin{figure}
\centering
\includegraphics{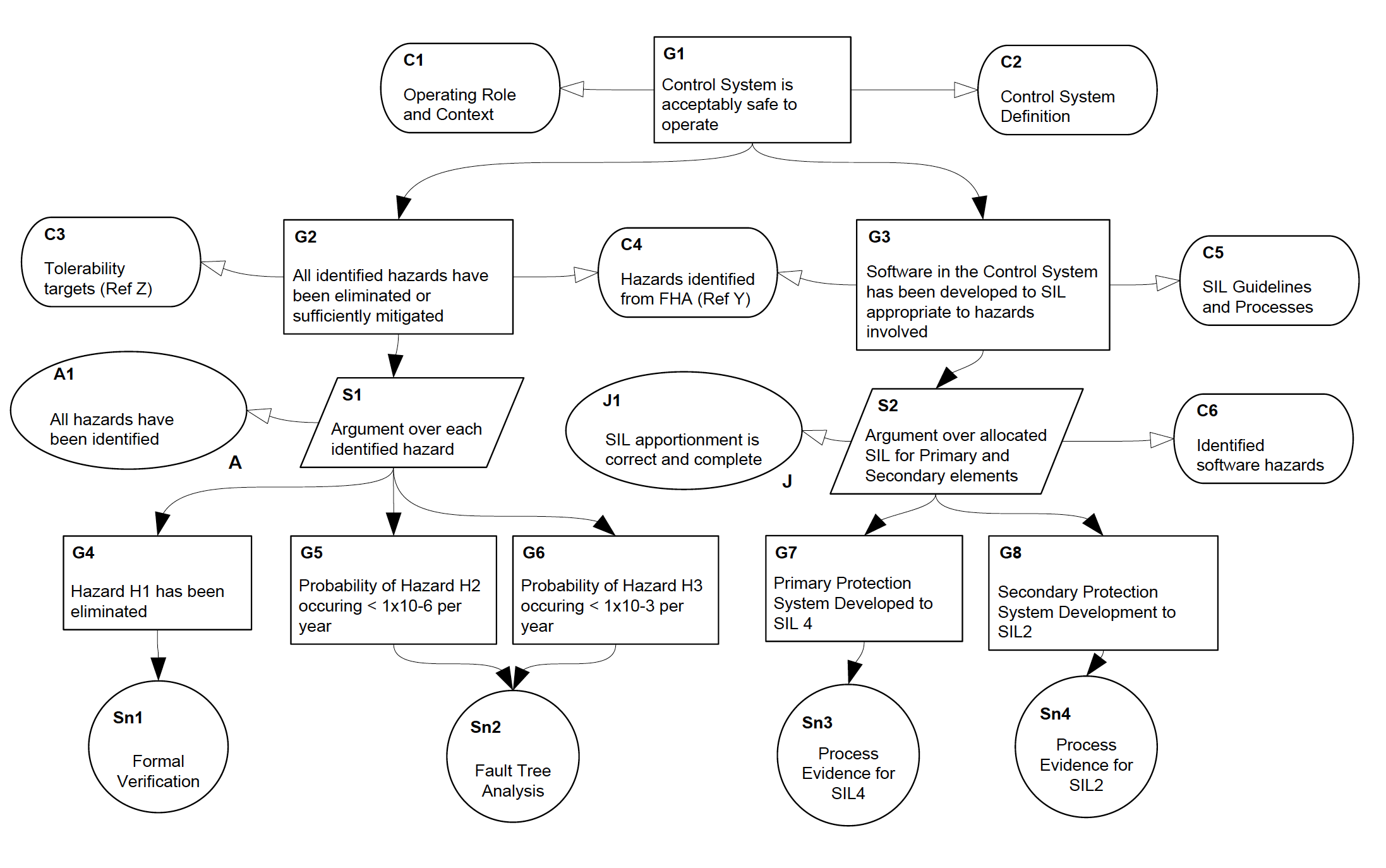}
\caption{An example assurance case focused on safety of a control system
(reprinted from \protect\hyperlink{ref-gsncommunity2018}{Community
2018})}
\end{figure}

\hypertarget{why-is-argument-based-assurance-useful}{%
\subsection{Why is argument-based assurance
useful?}\label{why-is-argument-based-assurance-useful}}

\begin{quote}
``Assurance cases are a primary means by which confidence in the safety
of the system is \emph{communicated to} and \emph{scrutinised by} the
diverse stakeholders, including regulators and policy-makers.''
(\protect\hyperlink{ref-habli2020}{Habli et al. 2020, 1}, emphasis
added)
\end{quote}

The above quotation emphasises the multi-purpose nature of ABA. For
instance, assurance cases,

\begin{itemize}
\tightlist
\item
  assist internal \textbf{reflection} and \textbf{deliberation} by
  providing a systematic and structured means for evaluating how the
  development of systems or products can fulfil certain normative goals
  (e.g.~safety or robustness), according to certain well-defined
  properties (e.g.~software hazards identified) and criteria (e.g.~risk
  reduction thresholds met)
\item
  provide a deliberate means for the \textbf{anticipation} and
  \textbf{pre-emption} of potential risks and adverse impacts through
  mechanisms of end-to-end assessment and redress;
\item
  facilitate \textbf{transparent communication} between developers and
  affected stakeholders;
\item
  support mechanisms and processes of \textbf{documentation} (or,
  reporting) to \textbf{ensure accountability} (e.g.~audits,
  compliance);
\item
  and build \textbf{trust and confidence} by promoting the adoption of
  best practices (e.g.~standards for warranted evidence) and by
  conveying the integration of these into design, development, and
  deployment lifecycles to impacted stakeholders.
\end{itemize}

The first two functions are particularly useful in industries or domains
where stringent regulation exists (e.g.~aviation, nuclear power).
However, as Cleland et al. (\protect\hyperlink{ref-cleland2012}{2012})
acknowledge, far from being a mere checklist exercise to ensure
compliance with regulatory requirements, the internal practice of
developing an assurance case can also offer organisations additional
insights into the \emph{sustainable maintenance} of business-critical
systems. For instance, as Habli et al.
(\protect\hyperlink{ref-habli2020}{2020}) note in the above quotation,
the communication of an assurance case is undertaken to enable
stakeholders---both within an organisation (e.g.~red teams) and outside
of the organisation (e.g.~auditors)---to \emph{critically scrutinise}
the evidence, argument, and assumptions that make up the overall case.
In doing so, gaps in the evidence base or weaknesses in an argument are
exposed, providing an opportunity to address any failures or concerns.

Over time, as best practices emerge, insights and practices derived from
the production of similar assurance cases can encourage the formation
and adoption of standards throughout an industry or domain. As such,
although assurance cases are a valuable \emph{communication tool}, which
help establish trust within and between organisations and wider
industries, the prerequisites for effective communication (e.g.~clarity,
accessible language) also serve to underwrite internal and external
practices of \emph{reflective} and \emph{anticipatory} governance
(e.g.~reflecting on possible challenges that may arise in the process of
development).

\hypertarget{building-an-assurance-case}{%
\subsection{Building an assurance
case}\label{building-an-assurance-case}}

\hypertarget{the-structure-elements-and-properties-of-an-assurance-case}{%
\subsubsection{The structure, elements, and properties of an assurance
case}\label{the-structure-elements-and-properties-of-an-assurance-case}}

The structure of an assurance case is a vital property that supports its
ability to \emph{provide assurance}. The structure of a particular case
may differ from others depending on the tools being used or the
framework being followed, and specific frameworks or tools may be better
suited to particular industries or contexts (e.g.~construction versus
healthcare). For example, in the Goal Structuring Notation (GSN)
framework---originally developed in the 1990s at the University of York
to assist the production and tracking of assurances in safety critical
industries such as traffic management and nuclear power---the principal
elements within an assurance case are goals, solutions, strategies,
contexts, assumptions, and justifications
(\protect\hyperlink{ref-gsncommunity2018}{Community 2018}).\footnote{The
  assurance case in Figure 1 uses GSN.}

Combining these elements into a \emph{structured} assurance case is a
systematic process, but can be streamlined by following methods such as
the following ``six-step'' approach
(\protect\hyperlink{ref-kelly1998}{Kelly 1998};
\protect\hyperlink{ref-hawkins2009}{R. D. Hawkins and Kelly 2009}):

\begin{itemize}
\tightlist
\item
  Step 1 - Identify goals (claims) to be supported
\item
  Step 2 - Define basis on which goals (claims) are stated
\item
  Step 3 - Identify strategy (argument approach) to support goals
  (claims)
\item
  Step 4 - Define basis on which the strategy (argument approach) is
  stated
\item
  Step 5 - Elaborate the strategy (argument)
\item
  Step 6 - Identify basic solution (evidence)
\end{itemize}

The process of starting with an approximate goal, and then explicating
its requirements to provide context for the development of an argument
and supporting evidence is why ABA can be viewed as a
\emph{goal-oriented} process.\footnote{Our own methodology will adopt a
  similar goal-oriented approach (see section
  \protect\hyperlink{ethical-assurance}{§4}).}

In addition to the structural features of an assurance case, there are
also important properties (or virtues) that support the purposes and
functions of assurance outlined in the previous section. For instance, a
good assurance case, among other things, should be \emph{clearly argued}
(e.g.~how does some evidence support a linked claim) and
\emph{accessible} (e.g.~language is plain, intelligible, and
non-technical where possible), in order to support \emph{transparent
communication}. It should also \emph{promote understanding and enquiry}
(e.g.~enables active enquiry from stakeholders) to help \emph{ensure
accountability}.

We can organise the desirable properties of an assurance case into the
following categories:

\begin{itemize}
\tightlist
\item
  Properties that support \textbf{reflection} and \textbf{deliberation}:
  systematic and standardised methodology, completeness of evidential
  support (e.g.~sufficient use of context/justification components).
\item
  Properties that support \textbf{documentation} and
  \textbf{accountability}: clearly identified and well-defined goals and
  claims, interoperability and reusability of argument patterns.
\item
  Properties that build \textbf{trust and confidence} through
  \textbf{transparent communication}: clear language and presentation,
  precision, accessibility, and simple, compelling and reasonable
  argument.
\end{itemize}

As a result of ABA's potential, there has been significant research into
the development of tools that can support and systematise the creation,
maintenance, assessment, and reporting of assurance cases
(\protect\hyperlink{ref-maksimov2018}{Maksimov et al. 2018}), in turn
helping to speed up the process of carrying out the aforementioned six
steps. For instance, there are tools that exist to support the building
of an assurance case by monitoring or tracking relevant artefacts that
are produced during the developmental process (e.g.~technical reports or
process logs), which can notify developers of relevant changes---in some
cases this may even result in dynamic and automatic changes to the
assurance cases (see \protect\hyperlink{ref-calinescu2018}{Calinescu et
al. 2018}). Another significant area of research, of relevance here, is
the development of \emph{argument patterns}.

\hypertarget{argument-patterns}{%
\subsubsection{Argument Patterns}\label{argument-patterns}}

\begin{quote}
``Argument patterns provide \emph{reusable templates} of the types of
claims, evidence, justification and contextual information that must be
covered in a compelling argument.''
(\protect\hyperlink{ref-picardi2020}{Picardi et al. 2020, 1}, emphasis
added)
\end{quote}

The creation (and adoption) of argument patterns supports a more
systematic (and efficient) approach to the development of assurance
cases, which can also support the adoption of industry-wide standards
and best practices as particular patterns become well-established.

Arguments that have been \emph{stress-tested} (i.e.~exposed to critical
scrutiny) and demonstrated to be successful in a previous setting can be
used as a basis to develop fundamental argument strategies that are
likely to work in similar settings (i.e.~constructing a generalisable
pattern). Patterns can include visual and formal representations and
templates of an argument's structure (e.g.~using GSN), but may also
incorporate additional prescriptive elements such as the \emph{intent}
behind the pattern, the \emph{applicability} or \emph{scope} of the
pattern, as well as any potential \emph{risks} or \emph{pitfalls} (see
\protect\hyperlink{ref-hawkins2011}{R. Hawkins et al. 2011}).

In addition to speeding up the process of creating an assurance case,
patterns can also function as useful deliberative prompts in an
anticipatory process that focuses attention on important tasks in the
design, development, or deployment of ML/AI systems. For instance, the
development of decision support systems in healthcare are likely to
require assurance that the outputs of the system effectively communicate
confidence estimates that are interpretable by the users
(e.g.~clinicians). As such, argument patterns can help generate
\emph{prospective} understanding for a project team (e.g.~anticipating a
possible weakness that limits the operational scope of the system)---in
a similar vein to the type of \emph{scenario planning} activities that
many organisations already undertake.

It is important to keep in mind that argument patterns are often
contextually-situated and domain-specific, however, due to the fact that
they are likely to have been developed \emph{bottom-up} by generalising
from a collection of similar assurance cases. For example, a pattern
that supports the creation of assurance cases focused on \emph{software
security} in financial systems will differ quite drastically from a
pattern to support the creation of assurance cases that support the
\emph{user safety} of medical devices. As such, although patterns can be
useful deliberative aids, it is important to consider whether specific
patterns are appropriate before using them, as there are likely to be
important limitations on the generalisability of most argument
patterns.\footnote{In the later sections we will return to this
  limitation when we discuss the evidential strength of assurance cases,
  which includes their diminishing reliability over time due to issues
  such as model drift that can occur in AI systems. It is well-known in
  argumentation theory that most forms of ABA rely on a process of
  either inductive or abductive reasoning. As such, the arguments
  themselves are typically \emph{defeasible}. However, as we will argue,
  this should be seen as an opportunity to promote genuine forms of
  inclusivity and democratic participation, rather than simply being a
  weakness of the methodology (see section
  \protect\hyperlink{conclusion-challenges-open-questions-and-next-steps}{5}).}

In the next section, we will begin the process of adapting ABA to
accommodate wider normative considerations, resulting in introduction of
the method of ethical assurance.

\hypertarget{from-safety-assurance-to-ethical-assurance}{%
\section{From Safety Assurance to Ethical
Assurance}\label{from-safety-assurance-to-ethical-assurance}}

In this section we start by addressing the limitations of existing
research on ML assurance, generally considered, explain \emph{why it
matters} that we address these gaps, and in turn motivate the need for
our positive proposal of ethical assurance that is introduced towards
the end of the section.

\hypertarget{existing-research-in-the-assurance-of-ml-systems}{%
\subsection{Existing Research in the Assurance of ML
Systems}\label{existing-research-in-the-assurance-of-ml-systems}}

ABA is not the only method of providing assurance. At present, there is
limited (but growing) research looking at the use of ABA for ML/AI
systems (see \protect\hyperlink{ref-ashmore2019}{Ashmore, Calinescu, and
Paterson 2019}; \protect\hyperlink{ref-picardi2020}{Picardi et al.
2020}; \protect\hyperlink{ref-ward2020}{Ward and Habli 2020} for some
notable exceptions). However, there is a wide variety of research that
supports the assurance of ML/AI systems more broadly construed. By this
we mean research that seeks to support or develop mechanisms by which
the processes and outcomes that characterise ML/AI research and
innovation can be made, for example, more transparent, trustworthy, or
responsible. Within this broader remit, there is research that includes
general overviews or frameworks that support transparent reporting and
communication (\protect\hyperlink{ref-mitchell2019}{Mitchell et al.
2019}; \protect\hyperlink{ref-brundage2020}{Brundage et al. 2020}),
specific (narrowly-focused) tools that support bias mitigation or
algorithmic interpretability
(\protect\hyperlink{ref-ibmresearch2018}{Research 2018};
\protect\hyperlink{ref-pair2020}{PAIR 2020};
\protect\hyperlink{ref-lundberg2020}{Lundberg 2020};
\protect\hyperlink{ref-ico2020b}{ICO 2020}), as well as more focused
extensions of assurance cases to address the specific challenges of ML
(\protect\hyperlink{ref-ashmore2019}{Ashmore, Calinescu, and Paterson
2019}; \protect\hyperlink{ref-ward2020}{Ward and Habli 2020};
\protect\hyperlink{ref-habli2020}{Habli et al. 2020}). Each of these
approaches can play a valuable role individually, but collectively add
up to a (currently) disorganised toolbox of practical mechanisms with
little unifying purpose or direction.

Brundage et al. (\protect\hyperlink{ref-brundage2020}{2020}) provide
some means for bringing order to this miscellany, by identifying
practical mechanisms to support the trustworthy development of AI
systems, categorised according to whether they are `institutional,'
`software,' or `hardware' mechanisms. For example, they note that
institutional mechanisms, such as \emph{third party auditing} can be
used to create a ``robust alternative to self-assessment claims,'' while
\emph{bias and safety bounties} can ``strengthen incentives to to
discover and report flaws in AI systems''
(\protect\hyperlink{ref-brundage2020}{Brundage et al. 2020, 1}).
Similarly, the software mechanisms they discuss also overlap with the
aforementioned tools that support tasks such as bias mitigation or
explainability. As such, their taxonomy could be used to support the
adoption of industry-wide standards. It also resonates with the recent
strategic directions of political institutions that emphasise the
importance of trustworthy AI development (e.g.
\protect\hyperlink{ref-high-levelexpertgrouponai2019}{High-Level Expert
Group on AI 2019}). However, insofar as the purpose of this taxonomy is
to help with the evaluation of ``verifiable claims,'' which the authors
define as ``statements for which evidence and arguments can be brought
to bear on the likelihood of those claims being true,'' the mechanisms
themselves are insufficient.\footnote{It is important to acknowledge
  that Brundage et al. (\protect\hyperlink{ref-brundage2020}{2020})
  recognise that the mechanisms alone are merely tools to support wider
  processes of governance, and also suggest the need for pursuing
  argument-based forms of assurance in Appendix III.}

Turning to arguably one of the most comprehensive proposals, Ashmore et
al. (\protect\hyperlink{ref-ashmore2019}{2019}) provide a systematic
survey of ML assurance, focusing on the generation of evidential
artefacts that can be used in the process of developing and evaluating
an ML system. In a similar vein to
(\protect\hyperlink{ref-brundage2020}{Brundage et al. 2020}), their
approach covers the methods and mechanisms that are capable of providing
evidence for claims about the properties of ML systems. However, their
approach categorises these methods according to \emph{where in the ML
lifecycle} they are most relevant. As they note, each of the stages
within this lifecycle have different \emph{desiderata} that affect the
generation of evidential artefacts. For example, they argue that (from
an assurance perspective) an ML model should exhibit the following
properties: performant, robust, reusable, and interpretable.\footnote{Ashmore
  et al. (\protect\hyperlink{ref-ashmore2019}{2019}) also define key
  desiderata for each of the four stages of their ``ML lifecyle'': data
  management, model learning, model verification, and model deployment.}
And, for each of these key desiderata, there are various methods that
can generate evidence to support corresponding claims (e.g., providing
details of regularization methods can verify claims about the robustness
of the model, and transfer learning can support reusability claims).

The survey of methods that Ashmore et al.
(\protect\hyperlink{ref-ashmore2019}{2019}) present is impressive, and
has more recently been extended by several of the original authors to
connect it more directly to ABA (see
\protect\hyperlink{ref-hawkins2021}{R. Hawkins et al. 2021}). The
\emph{process-based} strategy for representing the ML lifecycle employed
in this work is also adopted in our own proposal, with some notable
differences (see \protect\hyperlink{ethical-assurance}{§4}). As the
authors acknowledge (\protect\hyperlink{ref-ashmore2019}{Ashmore,
Calinescu, and Paterson 2019}), however, there are many open challenges
that arise as a result of the novel properties of ML systems, when
compared to traditional technological systems (e.g.~automated
decision-making capabilities).

To take just one example, Ward and Habli
(\protect\hyperlink{ref-ward2020}{2020}) develop an assurance case
pattern for the \emph{interpretability} of ML systems---a key desiderata
discussed in (\protect\hyperlink{ref-ashmore2019}{Ashmore, Calinescu,
and Paterson 2019}) and a vital component in recent efforts to improve
the explainability of AI systems (\protect\hyperlink{ref-ico2020b}{ICO
2020}). The pattern they propose is developed using the GSN notation
(i.e., a goal structured argument), and as such focuses on the assurance
of an \emph{interpretability claim} that takes on the following general
format:

\begin{quote}
The \{\emph{ML Model}\} is sufficiently \{\emph{interpretable}\} in the
intended \{\emph{context}\}.
\end{quote}

The variables, \{\emph{ML Model}\} and \{\emph{interpretable}\}, are
properly specified within the \{\emph{context}\} of a specific project,
allowing for variations in the type of interpretability considered to be
relevant (e.g.~inherent transparency of model, versus use of a post hoc
method for explaining how a particular outcome was obtained).
Interpretability is a vital property of systems that rely on ML models
or some form of AI and are used in safety-critical domains,
(e.g.~healthcare). Therefore, as Ward and Habli
(\protect\hyperlink{ref-ward2020}{2020}) argue, providing assurance that
a particular model is ``sufficiently interpretable'' in a given context
(i.e.~a particular time, setting, and a specified audience), helps build
confidence in the use of the system, and thus has ethical significance.
However, while the pattern that Ward and Habli develop is generalisable
to a range of contexts, it is nevertheless framed in terms of
\emph{safety concerns}.\footnote{For instance, consider the following
  statement from (\protect\hyperlink{ref-hawkins2021}{R. Hawkins et al.
  2021, 13}): ``requirements such as security or usability should be
  defined as ML safety requirements only if the behaviours or
  constraints captured by these requirements influence the safety
  criticality of the ML output. `Soft constraints' such as
  interpretability may be crucial to the acceptance of an ML component
  especially where the system is part of a socio‐technical solution. All
  such constraints defined as ML safety requirements must be clearly
  linked to safety outcomes.''} That is, the reason for assuring a
system's interpretability is grounded in the necessity of demonstrating
that it is `safe to operate.' While this has the effect of anchoring
requirements, such as interpretability, in clearly articulated safety
outcomes, it simultaneously divorces it from wider normative
considerations that are captured by more inclusive goals such as
explainability (\protect\hyperlink{ref-ico2020}{ICO and Institute
2020}), or related principles, such as respect for autonomy or informed
consent. Ward and Habli do acknowledge that the first step in the
process of developing an assurance case centred upon interpretability is
to ``ask why the project needs interpretability and set the desired
requirements that the project should satisfy.'' Therefore, it is
possible that the pattern they offer may also serve to provide assurance
for wider (interpretability-linked) normative goals.\footnote{We return
  to this point in \protect\hyperlink{top-level-normative-goal}{§4.3.1}
  when we discuss the relationship between multiple, complementary
  goals.}

These previous three examples are really just the tip of the iceberg in
this fast moving area of ML research and innovation. However, while
there are many valuable and worthwhile contributions, the existing
literature is nevertheless limited in several ways, all of which are
related to its \emph{scope}.

First, the proposed frameworks, which claim to be ``end-to-end,'' often
do not make sufficient room for \emph{wider normative considerations}
that arise \emph{prior to} stages such as data extraction or model
development. For instance, a growing literature has drawn attention to
the existence of historical or social biases that affect the fairness,
validity, performance, and trustworthiness of ML systems
(\protect\hyperlink{ref-binns2018}{Binns 2018};
\protect\hyperlink{ref-kalluri2020}{Kalluri 2020};
\protect\hyperlink{ref-benjamin2019}{Benjamin 2019};
\protect\hyperlink{ref-selbst2019}{Selbst et al. 2019}). The neglect of
such considerations (e.g.~the existence of historical patterns of social
discrimination) can often cause cascading effects through the ML
lifecycle, if not properly considered or addressed
(\protect\hyperlink{ref-leslie2021}{Leslie 2021}).

Second, the current literature is often too narrowly focused on
\emph{technological solutions} (e.g.~FairML ``solutions'' to complex
social justice issues). However, many social problems require a broader,
more nuanced, and deliberative approach, rarely reducible to a single
solution. Rather, as seen in the recent failure of many contact-tracing
systems, it is the absence of a legitimating \emph{social license} that
often leads to finite public resources going to waste
(\protect\hyperlink{ref-leslie2020a}{Leslie 2020}).

Finally, and perhaps most importantly for the present paper, the current
literature is focused on a \emph{limited set of goals}, such as safety
or security. While this is understandable in a technical industry
focused on regulatory compliance, the myopic focus on merely doing what
is \emph{necessary}, rather than what is \emph{best}, is often
unsustainable in the long-term. The market dynamics that may emerge from
such a collective attitude can generate a ``race to the bottom,'' as has
been seen recently in the domain of digital marketing and advertising
(e.g.~data privacy scandals).

These above limitations need to be addressed, if we are to fully embed
ethical considerations into the design, development, and deployment of
ML. Although ethical assurance, as a methodology, serves as a structured
means for responding to some of these limitations, it presupposes a
process-based form of ethical deliberation that serves as an anchor for
the reflective activities that ought to be undertaken by project teams
in the service of developing an ethical assurance case. Therefore, we
turn now to a brief description of this process.

\hypertarget{foundational-values}{%
\subsection{Foundational Values}\label{foundational-values}}

\hypertarget{operationalising-normative-principles}{%
\subsection{Operationalising Normative
Principles}\label{operationalising-normative-principles}}

There is a lot of confusion about the role that normative principles
play in supporting ethical reasoning and decision-making. A common set
of worries hinge upon the misconception that ethical principles such as
\emph{respect for autonomy} or \emph{fairness} are too vague or abstract
to be \emph{practically useful} in a technical context (see
\protect\hyperlink{setting-the-stage}{§1}), can disguise deep-seated
disagreement, and even enable ethics washing. These worries are borne,
however, from a misunderstanding of the role that principles are
designed to play in ethical deliberation. In brief, while ethical
principles are not \emph{sufficient} in themselves for action-guidance,
they do play a vital, contributory (and sometimes explanatory or
justificatory) role in deliberation. In this section, we argue that they
are best viewed as \emph{starting points} in an ongoing, participatory
process of reflection, action, and justification.\footnote{This view
  derives, primarily, from the ethical theory of principlism
  (\protect\hyperlink{ref-beauchamp2013}{Tom L. Beauchamp and Childress
  2013}), but is also reflected in contemporary research in responsible
  research and innovation (\protect\hyperlink{ref-owen2013}{Owen,
  Bessant, and Heintz 2013}).}

To begin with, ethical principles can support a practical, reflective
process of anticipation during the planning stages of a project by
offering foundations or starting points for identifying and evaluating
the various ethical decision points that will invariably arise during a
project's lifecycle. They are intended to be \emph{starting points} for
reflection, however, due to the fact that the linked concepts
(e.g.~`fairness') may a) mean different things to different people, and
b) demand very different actions in the context of different projects.
For example, the principle of `respect for individual autonomy' will
mean very different things for two projects that aim to develop an AI
chatbot, where one supports the education of children and the other is
used to support the assessment of individuals suffering from mental
health disorders. That is, the principle is \emph{specified} in
different ways for the two projects, despite serving as a fundamental
standard of conduct from which other related moral standards or
judgement are drawn. For instance, a healthcare professional may
recognise that restricting an individual's freedom, if they are
suffering from a severe mental health disorder, is necessary, while
nevertheless doing so in a way that respects their individual autonomy
during care and treatment. In this way, ethical principles may be
binding, but only as \emph{pro tanto} reasons for action---that is,
reasons that speak in favour of some goal or claim but that can be
overridden by additional, competing reasons. Because of this, it is
important to use a process of critical and inclusive reflection not only
to identify the relevant principles that underpin a project, but also to
identify the different interpretations that affected stakeholders may
have regarding the principles in question.

This point speaks to the (admittedly misplaced) confusion that is
directed towards the role of ethical principles in the domain of
technology governance. General principles do not \emph{fully determine}
actions or judgements; their substantive content is insufficient for
directing action without first addressing how the principle is specified
in a particular context (e.g.~the principle of `transparency' in the
context of criminal justice or healthcare where disclosure of sensitive
information ought to be restricted). This requires additional reflection
and deliberation, often with domain experts and stakeholders. Hence,
principles should be treated, as Beauchamp and Childress
(\protect\hyperlink{ref-beauchamp2004}{2004}) note, ``less as firm
directives that are applied and more as general guidelines that are
explicated and made suitable for specific tasks, as often occurs in
formulating policies and altering practices.''

Because of this, principles can be supported and complemented by
additional processes (e.g.~the processes of ethical assurance that we
set out in the next section). Well-established virtues, for example, can
help ensure that any actions taken throughout a project are in alignment
with the general guidance of relevant principles (e.g.~choosing to use
an automated monitoring tool to assess a quantifiable notion of
fairness). In addition, practical mechanisms such as well-designed
stakeholder impact assessments can ensure that possible risks and harms
to different groups are identified early on. And, regulatory guidance or
legal precedents can \emph{constrain} decision-making in positive ways,
rather than merely \emph{restricting} or \emph{inhibiting} innovation.

This understanding of the \emph{complementary} and
\emph{socially-embedded} role of ethical principles as supporting and
guiding practical decision-making is sometimes known as the ``reflective
equilibrium'' model (\protect\hyperlink{ref-rawls1999}{Rawls
1999})---referring to the stable point of a deliberative process in
which a group of individuals (or society more broadly) reflect on and
revise a moral belief or judgement. The resulting judgement aims at
maximising coherence among the linked set of beliefs that underpin the
judgement, such that the holding of the judgement, while defeasible, is
nevertheless justified (or, warranted) given the deliberative process.
Of course, to be legitimate in the first place, this process of
deliberation has to adhere to additional standards that are determined
by the communicative context and demanded as normative preconditions of
discursive will formation (e.g.~following a \emph{rational procedure};
representing a fair \emph{democratic process})
(\protect\hyperlink{ref-beauchamp2004}{Tom L. Beauchamp and DeGrazia
2004}; \protect\hyperlink{ref-habermas1998}{Habermas 1998}).

These previous discussion regarding the role of ethical principles
belies an enormous complexity, which is well beyond the scope of this
paper to explore. In closing, we wish to highlight a few salient
details. Even if these points help to address some of the confusion
about the role that ethical values and principles are \emph{intended} to
play, some practitioners may remain unsatisfied. For instance, as
(\protect\hyperlink{ref-raji2020}{Raji et al. 2020, 33}) note,
``{[}\ldots{]} it remains challenging for practitioners to identify the
harmful repercussions of their own systems prior to deployment, and,
once deployed, emergent issues can become difficult or impossible to
trace back to their source.'' This challenge is sometimes referred to as
Collingridge's dilemma, named after David Collingridge
(\protect\hyperlink{ref-collingridge1980}{Collingridge 1980}), and is
well-known in the area of responsible research and innovation
(\protect\hyperlink{ref-owen2013}{Owen, Bessant, and Heintz 2013}).

Ultimately, a process of ongoing reflection can help \emph{mitigate} the
risk of harms associated with systems prior to their deployment. For
instance, ensuring that specific harms to marginalised groups are
identified through an inclusive and participatory process of stakeholder
engagement to ensure that easily identifiable issues do not go
unaddressed. But, in spite of this, a degree of epistemic humility is
invariably required, as some risks may be impossible to remove entirely,
resulting in unintended consequences---especially with novel
technologies that shape our sociotechnical environments in unforeseeable
ways.\footnote{This does not mean that if a system has been deployed
  with little to no oversight, nor with any due consideration given to
  the transparency and accountability of the processes, and ends up
  causing significant harm, that those responsible should be able to
  claim that it was due to ``unforeseeable risk.''} What is important
here is that practical measures are taken to promote \emph{best
practices}. This means building in space for reflection
\emph{throughout} the processes of design, development, and deployment,
and building in time to return to previous steps if issues are only
identified at a later stage in the project lifecycle. In the next
section, we will discuss how the process of ethical assurance
accommodates this requirement throughout the lifecycle of a project.

\hypertarget{ethical-assurance}{%
\section{Ethical Assurance}\label{ethical-assurance}}

In this section we present the `ethical assurance' methodology. The
section is structured as follows:

\begin{itemize}
\tightlist
\item
  First we explain how ethical assurance differs from ABA and how it
  extends the original method.
\item
  Second we introduce a heuristic model of a typical project lifecycle
  for an ML/AI system, which is used to ground the ethical assurance
  methodology.
\item
  Third, we show how the methodology is grounded in this model,
  introducing the process of reflect, act, and justify as a way of
  providing a scaffold for the building of an ethical assurance case.
\item
  Fourth, we discuss each of the elements of an ethical assurance
  case---top-level normative goal, system or project property claim,
  argument (warrant), evidential claim, and evidence---and explain how
  they are connected.
\item
  Finally, we explore some ancillary considerations for ethical
  assurance, such as how to evaluate evidence, and how to generalise
  ethical assurance cases to develop argument patterns.
\end{itemize}

\hypertarget{what-is-ethical-assurance}{%
\subsection{What is ethical
assurance?}\label{what-is-ethical-assurance}}

\begin{quote}
``an oral or written expression is a standpoint if it expresses a
certain positive or negative position with respect to a proposition,
thereby making it plain what the speaker or writer stands for. And a
series of utterances constitutes an argumentation only if these
expressions are jointly used in an attempt to justify or refute a
proposition, meaning that they can be seen as a concerted effort to
defend a standpoint in such a way that the other party is convinced of
its acceptability.'' (\protect\hyperlink{ref-eemeren2004}{Eemeren and
Grootendorst 2004, 3})
\end{quote}

The above quotation from van Eemeren and Grootendorst
(\protect\hyperlink{ref-eemeren2004}{2004}) sets an important anchoring
point for the following section, by making explicit that an ethical
assurance case is an argument that seeks to justify a \emph{particular
standpoint} towards some proposition or claim (i.e.~the ethical goal).
The importance of this feature will become clearer as the section
progresses and we introduce the structure and elements of ethical
assurance.

Ethical assurance is a version of ABA that retains much of the original
approach but extends it to incorporate wider ethical concerns into the
design, development, and deployment of data-driven technologies, such as
ML/AI, in a systematic and pragmatic manner. This is reflected in the
following definition, which retains much of our earlier definition of
ABA:

\begin{quote}
Ethical assurance is a process of using structured argumentation to
provide reviewable assurance that a particular set of normative claims
about the corresponding ethical properties of a system are warranted
given the available evidence.
\end{quote}

In spite of the similarities, an extension of ABA is needed because the
traditional focus of assurance cases (e.g.~on safety and reliability),
while important and vital, is necessarily limited due to the alternative
goals (e.g.~compliance with technical standards). For instance, consider
how a power station can operate safely and reliably in certain respects,
while nevertheless harming the environment through pollution
(\protect\hyperlink{ref-raji2020}{Raji et al. 2020})\footnote{Harm to
  the environment can of course be incorporated into a broader notion of
  `safety,' such that pollution generated in the everyday operations of
  a power station are factored into a safety assessment. However, the
  point we wish to address here is that the scope of concepts, such as
  `safety' and `reliability' tends to reflect a domain-specific focus or
  set of priorities (e.g.~compliance with technical or legal standards,
  rather than ethical principles).}. Furthermore, as the explicit
inclusion of \emph{reviewability} suggests, ethical assurance supports
and promotes a culture of active enquiry, necessary for ensuring the
moral legitimacy of a project
(\protect\hyperlink{ref-oneill2002}{O'Neill 2002}).

Moreover, this extension to accommodate wider ethical concerns also
incorporates additional steps that a) identify relevant ethical values
that are implicated in the design, development, and deployment of an AI
technology, and b) support the evaluation and operationalisation of
ethical principles that serve as important guardrails for practical
decision-making. Many of these additional steps are anticipatory in
nature, insofar as they require a project team to reflect on the
normative properties of their project and system, and determine which
actions need to be taken to ensure that any related benefits are
realised and any risks of harm are mitigated. This preliminary,
reflective work is where ethical assurance departs most, though not
entirely, from traditional forms of ABA.

To help distinguish this aspect of ethical assurance we will refer to
the procedure of building an ethical assurance case as a process of
\emph{reflection}, \emph{action}, and \emph{justification}. This section
will expand on each of these stages in further detail, connecting them
to the elements of an ethical assurance case. However, before showing
how they are connected to the key elements of an assurance case, it is
necessary to show how reflection, action, and justification is grounded
in a practical process of design, development, and deployment.

\hypertarget{a-sociotechnical-approach-to-the-mlai-project-lifecycle}{%
\subsection{A Sociotechnical Approach to the ML/AI Project
Lifecycle}\label{a-sociotechnical-approach-to-the-mlai-project-lifecycle}}

There are many ways of carving up a project lifecycle for a ML/AI system
or other data-driven technology (hereafter shortened to just `project
lifecycle'). For instance, Sweenor et al.
(\protect\hyperlink{ref-sweenor2020}{2020}) break it into four stages:
Build, Manage, Deploy \& Integrate, Monitor.\footnote{These four stages
  are influenced by an `ML OPs' perspective
  (\protect\hyperlink{ref-sweenor2020}{Sweenor et al. 2020}). The term
  `MLOps' refers to the application of DevOps practices to ML pipelines.
  The term is often used in an inclusive manner to incorporate
  traditional statistical or data science practices that support the ML
  lifecycle, but are not themselves constitutive of machine learning
  (e.g.~exploratory data analysis), as well as deployment practices that
  are important within business and operational contexts
  (e.g.~monitoring KPIs).} Similarly,
(\protect\hyperlink{ref-ashmore2019}{Ashmore, Calinescu, and Paterson
2019}) identify four stages, which have a more specific focus on data
science: data management, model learning, model verification, and model
deployment. Furthermore, there are also well-established methods that
seek to govern common tasks within a project lifecyle, such as data
mining (e.g.~CRISP-DM or SEMMA).

The multiplicity of approaches is likely a product of the evolution of
diverse methods in data mining/analytics, the significant impact of ML
on research and innovation, and the specific practices and
considerations inherent to each of the various domains where ML
techniques are applied. While there are many benefits of existing
frameworks (e.g.~carving up a complex process into smaller components
that can be managed by a network of teams or organisations), they do not
tend to focus on the wider social or ethical aspects that interweave
throughout the various stages of a ML lifecycle. Figure 2, therefore,
presents a model of the ML lifecycle, which we have designed to support
the process of building an ethical assurance case, while remaining
faithful to the importance of technical requirements and challenges and
also supporting more open, reflective, and participatory forms of
deliberation.

\begin{figure}
\centering
\includegraphics{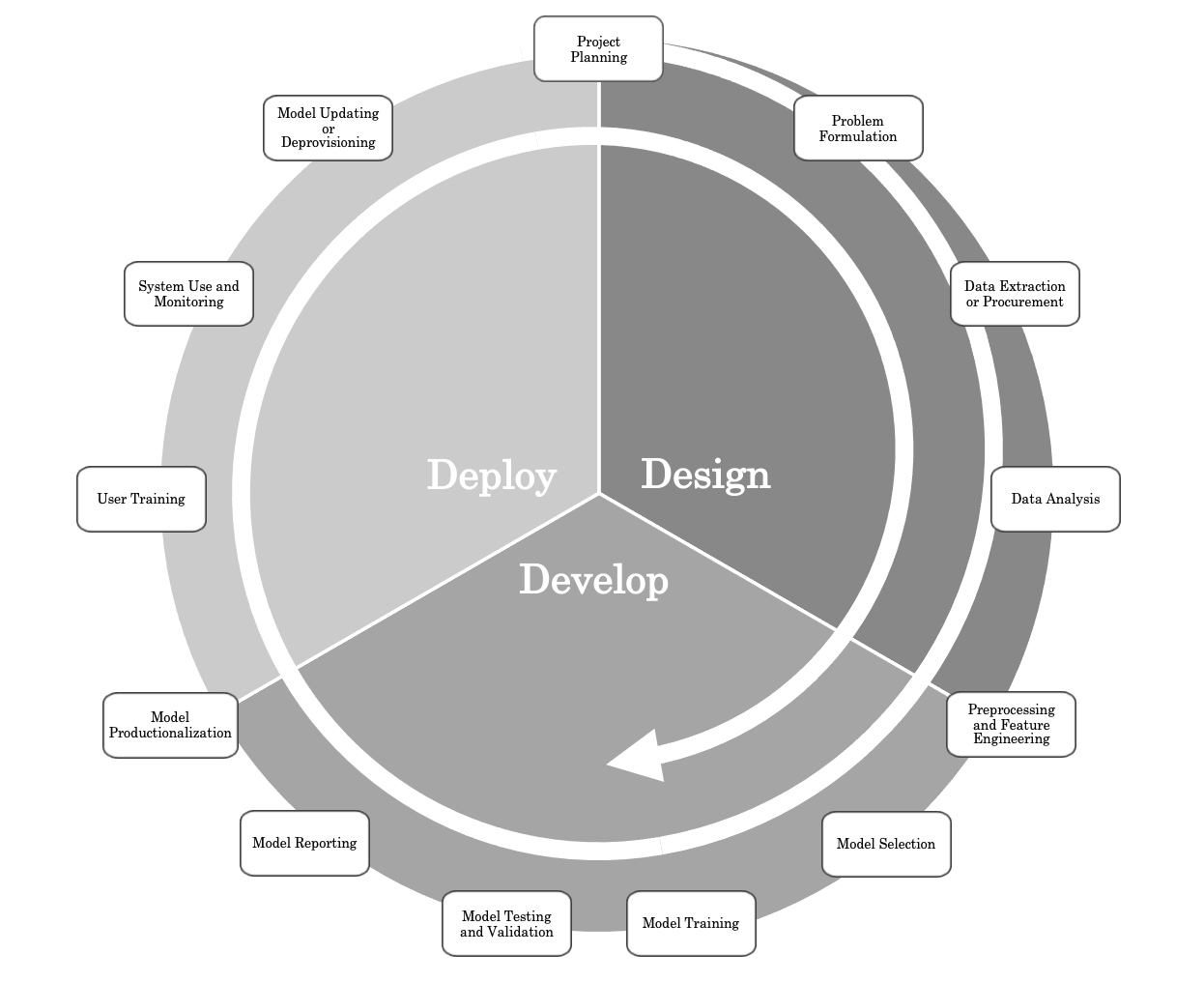}
\caption{The Project Lifecycle---the overarching stages of design,
development, and deployment (for a typical data-driven project) can be
split into indicative tasks and activities. In practice, both the stages
and the tasks will overlap with their neighbours, and may be revisited
where a particular task requires an iterative approach. The spiral
indicates that this is a diachronic, macroscopic process that evolves
and develops over time, and as the deployment stage finishes, a new
iteration is likely to begin.}
\end{figure}

To begin, the inner circle breaks the project lifecycle into a process
of (project) design, (model) development, and (system) deployment. These
terms are intended to be maximally inclusive. For example, the
\emph{design stage} encompasses any project task or decision-making
process that scaffolds or sets constraints on later project stages
(i.e.~design system constraints). Importantly, this includes ethical,
social, and legal constraints, which we will discuss later.

Each of the stages shades into its neighbours, as there is no clearly
delineated boundary that differentiates certain \emph{project} design
activities (e.g.~data extraction and exploratory analysis) from
\emph{model} design activities (e.g.~preprocessing and feature
engineering, model selection). As such, the design stage overlaps with
the \emph{development stage}, but the latter extends to include the
actual process of training, testing, and validating a ML model.
Similarly, the process of productionalizing a model within its runtime
environment can be thought of as both a development and deployment
activity. And, so, the \emph{deployment stage} overlaps with the
`development' stage, and also overlaps with the `design' stage, as the
deployment of a system should be thought of as an ongoing process
(e.g.~where new data are used to continuously train the ML model, or,
the decision to de-provision a model may require the planning and design
of a new model, if the older (legacy) system becomes outdated). For
these reasons, the project lifecycle is depicted as a spiral. However,
despite the unidirectional nature of the arrows, we also acknowledge
that ML/AI research and innovation is frequently an iterative process.
Therefore, the singular direction is only present at a macroscopic level
of abstraction (i.e., the \emph{overall} direction of progress for a
project), and allows for some inevitable back and forth between the
stages at the microscopic level.

The three higher-level stages can be thought of as a useful heuristic
for approaching the project lifecycle. However, each higher-level stage
subsumes a wide variety of tasks and activities that are likely to be
carried out by different individuals, teams, and organisations,
depending on their specific roles and responsibilities (e.g.~procurement
of data). Therefore, it is important to break each of the three
higher-level stages into their (typical) constituent parts, which are
likely to vary to some extent between specific projects or within
particular organisations. In doing so, we expose a wide range of diverse
tasks, each of which give rise to a variety of ethical, social, and
legal challenges. The following sections provides an illustrative
overview of these stages and tasks, as well as a \emph{non-exhaustive}
sample of the associated challenges.

\hypertarget{project-design-tasks-and-processes}{%
\subsubsection{(Project) Design Tasks and
Processes}\label{project-design-tasks-and-processes}}

\textsc{Project Planning}

Rather than using ML/AI as a ``hammer'' to go looking for nails, it is
best to have a clear idea in mind of what the project's goals are at the
outset. This can help to avoid a myopic focus on a narrow class of
ML/AI-based ``solutions,'' and also helps create space for a diversity
of approaches---some of which may not require ML/AI at all. Project
planning, therefore, can comprise a wide variety of tasks, including,
but not limited to:

\begin{itemize}
\tightlist
\item
  an assessment of whether building an AI model is the right approach
  given available resources and data, existing technologies and
  processes already in place, the complexity of the use-contexts
  involved, and the nature of the policy or social problem that needs to
  be solved (Leslie et al 2021a);
\item
  an analysis of user needs in relation to the prospective AI model and
  whether a solution involving the latter provides appropriate
  affordances in keeping with user needs and related functional
  desiderata;
\item
  mapping of key stages in the project to support governance and
  business tasks (e.g.~scenario planning);
\item
  an assessment of resources and capabilities within a team, which is
  necessary for identifying any skills gaps,
\item
  a contextual assessment of the target domain and of the expectations,
  norms, and requirements that derive therefrom;
\item
  stakeholder analysis and team positionality reflection to determine
  the appropriate level and scope of community engagement activities
  (Leslie et al 2021b);
\item
  stakeholder impact assessment, supported by affected people and
  communities, to identify and evaluate possible harms and benefits
  associated with the project (e.g.~socioeconomic inequalities that may
  be exacerbated as a result of carrying out the project), to gain
  social license and public trust, and also feed into the process of
  problem formulation in the next stage;
\item
  wider impact assessments---both where required by statute and done
  voluntarily for transparency and best practice (e.g.~equality impact
  assessments, data protection impact assessments, human rights impact
  assessment, bias assessment)
\end{itemize}

\textsc{Problem Formulation}

Here, `problem' refers both to a well-defined computational process (or
a higher-level abstraction of the process) that is carried out by the
algorithm to map inputs to outputs and to the wider practical, social,
or policy issue that will be addressed through the translation of that
issue into the statistical or mathematical frame. For instance, on the
computational side, a convolutional neural network carries out a series
of successive transformations by taking (as input) an image, encoded as
an array, in order to produce (as output) a decision about whether some
object is present in the image. On the practical, social, and policy
side, there will be a need to define the computational ``problem'' being
solved in terms of the algorithmic system's embeddedness in the social
environment and to explain how it contributes to (or affects) the wider
sociotechnical issue being considered. In the convolutional neural
network example, the system being produced may be a facial recognition
technology that responds to a perceived need for the biometric
identification of criminal suspects by matching face images in a police
database. The social issue of wanting to identify suspects is, in this
case, translated into the computational mechanism of the computer vision
system. But, beyond this, diligent consideration of the practical,
social, or policy issue being addressed by the system will also trigger,
\emph{inter alia}, reflection on the complex intersection of potential
algorithmic bias, the cascading effects of sociohistorical patterns of
racism and discrimination, wider societal and community impacts, and the
potential effects of the use of the model on the actors in the criminal
justice systems who will become implementers and subjects of the
technology.

Sociotechnical considerations are also important for determining and
evaluating the choice of target variables used by the algorithm, which
may ultimately be implemented within a larger automated decision-making
system (e.g.~in a verification system). The task of formulating the
problem allows the project team to get clear on what input data will be
needed, for what purpose, and whether there exists any representational
issues in, for example, how the target variables are defined. It also
allows for a project team (and impacted stakeholders) to reflect on the
reasonableness of the measurable proxy that is used as a mathematical
expression of the target variable, for instance, whether being taken
into care within six months of a visit from child protective services is
a reasonable proxy for a child's being ``at risk'' in a predictive risk
model for children's social care. The semantic openness and
contestability of formulating problems and defining target variables in
ML/AI innovation lifecycles is why stakeholder engagement, which helps
bring a diversity of perspectives to project design, is so vital, and
why this stage is so closely connected with the interpretive burdens of
the project planning stage (e.g.~discussion about legal and ethical
concerns regarding permissible uses of personal or sensitive
information).

\textsc{Data Extraction or Procurement}

Ideally, the project team should have a clear idea in mind (from the
planning and problem formulation stages) of what data are needed prior
to extracting or procuring them. This can help mitigate risks associated
with over-collection of data (e.g.~increased privacy or security
concerns) and help align the project with values such as \emph{data
minimisation} (\protect\hyperlink{ref-ico2020}{ICO and Institute 2020}).
Of course, this stage may need to be revisited after carrying out
subsequent tasks (e.g.~preprocessing, model testing) if it is clear that
insufficient or imbalanced data were collected to achieve the project's
goals. Where data is procured, questions about provenance arise
(e.g.~legal issues, concerns about informed consent of human data
subjects). Generally, responsible data extraction and procurement
require the incorporation of domain expertise into decision-making so
that desiderata of data minimisation as well as of securing relevant and
sufficient data can be integrated into design choices.

\textsc{Data Analysis}

Exploratory data analysis is an important stage for hypothesis
generation or uncovering possible limitations of the dataset that can
arise from missing data, in turn identifying the need for any subsequent
augmentation of the dataset to deal with possible class imbalances.
However, there are also risks that stem from cognitive biases
(e.g.~confirmation bias) that can create cascading effects that effect
downstream tasks (e.g.~model reporting).

\hypertarget{model-development-tasks-and-processes}{%
\subsubsection{(Model) Development Tasks and
Processes}\label{model-development-tasks-and-processes}}

\textsc{Preprocessing and Feature Engineering}

Pre-processing and feature engineering is a vital but often lengthy
process, which overlaps with the design tasks in the previous section
and shares with them the potential for human choices to introduce biases
and discriminatory patterns into the ML/AI workflow. Tasks at this stage
include data cleaning, data wrangling or normalisation, and data
reduction or augmentation. It is well understood that the methods
employed for each of these tasks can have a significant impact on the
model's performance (e.g.~deletion of rows versus imputation methods for
handling missing data). As Ashmore et al.
(\protect\hyperlink{ref-ashmore2019}{2019}) note, there are also various
desiderata that motivate the tasks, such as the need to ensure the
dataset that will feed into the subsequent stages is relevant, complete,
balanced, and accurate. At this stage, human decisions about how to
group or disaggregate input features (e.g.~how to carve up categories of
gender or ethnic groups) or about which input features to exclude
altogether (e.g.~leaving out deprivation indicators in a predictive
model for clinical diagnostics) can have significant downstream
influences on the fairness and equity of an ML/AI system.

\textsc{Model Selection}

This stage determines the model type and structure that will be produced
in the next stages. In some projects, model selection will result in
multiple models for the purpose of comparison based on some performance
metric (e.g.~accuracy). In other projects, there may be a need to first
of all implement a pre-existing set of formal models into code. The
class of relevant models is likely to have been highly constrained by
many of the previous stages (e.g.~available resources and skills,
problem formulation), for instance, where the problem demands a
supervised learning algorithm instead of an unsupervised learning
algorithm; or where explainability considerations require a more
interpretable model (e.g.~a decision tree).

\textsc{Model Training}

Prior to training the model, the dataset will need to be split into
training and testing sets to avoid model overfitting. The \emph{training
set} is used to fit the ML model, whereas the \emph{testing set} is a
hold-out sample that is used to evaluate the fit of the ML model to the
underlying data distribution. There are various methods for splitting a
dataset into these components, which are widely available in popular
package libraries (e.g.~the scikit-learn library for the Python
programming language). Again, human decision-making at this stage about
the training-testing split and about how this shapes desiderata for
external validation---a subsequent process where the model is validated
in wholly new environments---can be very consequential for the
trustworthiness and reasonableness of the development phase of an ML/AI
system.

\textsc{Model Validation and Testing}

The testing set is typically kept separate from the training set, in
order to provide an unbiased evaluation of the final model fit on the
training dataset. However, the training set can be further split to
create a validation set, which can then be used to evaluate the model
while also \emph{tuning model hyperparameters}. This process can be
performed repeatedly, in a technique known as (k-fold) cross-validation,
where the training data are resampled (\emph{k}-times) to compare models
and estimate their performance in general when used to make predictions
on unseen data. This type of validation is also known as `internal
validation,' to distinguish it from external validation, and, in a
similar way to choices made about the training-testing split, the manner
in which it is approached can have critical consequences for how the
performance of a system is measured against the real-world conditions
that it will face when operating ``in the wild.''

\textsc{Model Reporting}

Although the previous stages are likely to create a series of artefacts
while undertaking the tasks themselves, model reporting should also be
handled as a separate stage to ensure that the project team reflect on
the future needs of various stakeholders and end users. While this stage
is likely to include information about the performance measures used for
evaluating the model (e.g.~decision thresholds for classifiers, accuracy
metrics), it can (and should) include wider considerations, such as
intended use of the model, details of the features used,
training-testing distributions, and any ethical considerations that
arise from these decisions (e.g.~fairness constraints, use of
politically sensitive demographic features).\footnote{There is some
  notable overlap between this stage of the project lifecyle and the
  ethical assurance methodology, as some approaches to model reporting
  often contain similar information that is used in building an ethical
  assurance case (\protect\hyperlink{ref-mitchell2019}{Mitchell et al.
  2019}; \protect\hyperlink{ref-ashmore2019}{Ashmore, Calinescu, and
  Paterson 2019}), specifically in the process of establishing
  evidential claims and warrant (see
  \protect\hyperlink{evidential-claim-and-artefact}{§§4.3.3-4.3.4}).}

\hypertarget{system-deployment-tasks-and-processes}{%
\subsubsection{(System) Deployment Tasks and
Processes}\label{system-deployment-tasks-and-processes}}

\textsc{Model Productionalization}

Unless the end result of the project is the model itself, which is
perhaps more common in scientific research, it is likely that the model
will need to be implemented within a larger system. This process,
sometimes known as `model operationalisation,' requires understanding
(a) how the model is intended to function in the proximate system
(e.g.~within an agricultural decision support system used to predict
crop yield and quality) and (b) how the model will impact---and be
impacted by---the functioning of the wider sociotechnical environment
that the tool is embedded within (e.g.~a decision support tool used in
healthcare for patient triaging that may exacerbate existing health
inequalities within the wider community). Ensuring the model works
within the proximate system can be a complex programming and software
engineering task, especially if it is expected that the model will be
updated continuously in its runtime environment. But, more importantly,
understanding how to ensure the model's sustainability given its
embeddedness in complex and changing sociotechnical environments
requires active and contextually-informed monitoring, situational
awareness, and vigilant responsiveness.

\textsc{User Training}

Although the performance of the model is evaluated in earlier stages,
the model's impact cannot be entirely evaluated without consideration of
the human factors that affect its performance in real-world settings.
The impact of human cognitive biases, such as algorithmic
aversion\footnote{Algorithmic aversion refers to the reluctance of human
  agents to incorporate algorithmic tools as part of their
  decision-making processes due to misaligned expectations of the
  algorithm's performance (see \protect\hyperlink{ref-burton2020}{Burton
  et al. 2020}).} must also be considered, as such biases can lead to
over- and under-reliance on the model (or system), in turn negating any
potential benefits that may arise from its use. Understanding the social
and environmental context is also vital, as sociocultural norms may
contribute to how training is received, and how the system itself is
evaluated (see \protect\hyperlink{ref-burton2020}{Burton et al. 2020}).

\textsc{System Use and Monitoring}

Depending on the context of deployment, it is likely that the
performance of the model could degrade over time. This process of
\emph{model drift} is typically caused by increasing variation between
how representative the training dataset was at the time of development
and how representative it is at later stages, perhaps due to changing
social norms (e.g.~changing patterns of consumer spending, evolving
linguistic norms that affect word embeddings). As such, mechanisms for
monitoring the model's performance should be instantiated within the
system's runtime protocols to track model drift, and key thresholds
should be determined at early stages of a project (e.g.~during project
planning or in initial impact assessment) and revised as necessary based
on monitoring of the system's use.

\textsc{Model Updating or De-provisioning}

As noted previously, model updating can occur continuously if the
architecture of the system and context of its use allows for it.
Otherwise, updating the model may require either revisiting previous
stages to make planned adjustments (e.g.~model selection and training),
or if more significant alterations are required the extant model may
need to be entirely de-provisioned, necessitating a return to a new
round of project planning.

This overview and summary of the project lifecycle is by necessity an
abstraction. However, it provides a useful anchor for subsequent
discussion, and serves to motivate the following question: how do you
provide assurance for the diversity of tasks included throughout the
process? For instance, there may be a plurality of ethical goals
relevant to the assurance of \textsc{(model) development} or
\textsc{system use and monitoring}, including demonstrating that the
system being deployed is safe, secure, fair, trustworthy, explainable,
sustainable, or respectful of human agency and autonomy. How do you
provide assurance that the interconnected project processes and
activities individually and collectively support the relevant goal? This
is why we need a unifying framework and methodology that makes space for
the operationalisation of end-to-end, normative considerations and
complements existing regulatory culture, as opposed to merely a
miscellany of practical mechanisms.

\hypertarget{the-structure-and-elements-of-an-ethical-assurance-case}{%
\subsection{The structure and elements of an ethical assurance
case}\label{the-structure-and-elements-of-an-ethical-assurance-case}}

The general structure of ethical assurance, or the building of an
ethical assurance case more specifically, can be described as an
iterative and cyclical process of reflection, action, and justification
throughout the stages of the project lifecycle that were outlined in the
previous sub-section. This iterative process involves (a) establishing
the normative goals that identify and articulate key ethical qualities
and determining the properties needed to assure these goals, (b) taking
actions to operationalise these properties, and (c) compiling the
evidence of these actions that then provides warrant for claims that the
goals have been ascertained. The process of developing an assurance case
assists internal reflection and deliberation, promoting the adoption of
best practices and integrating these into design, development, and
deployment lifecycles.

\begin{figure}
\centering
\includegraphics{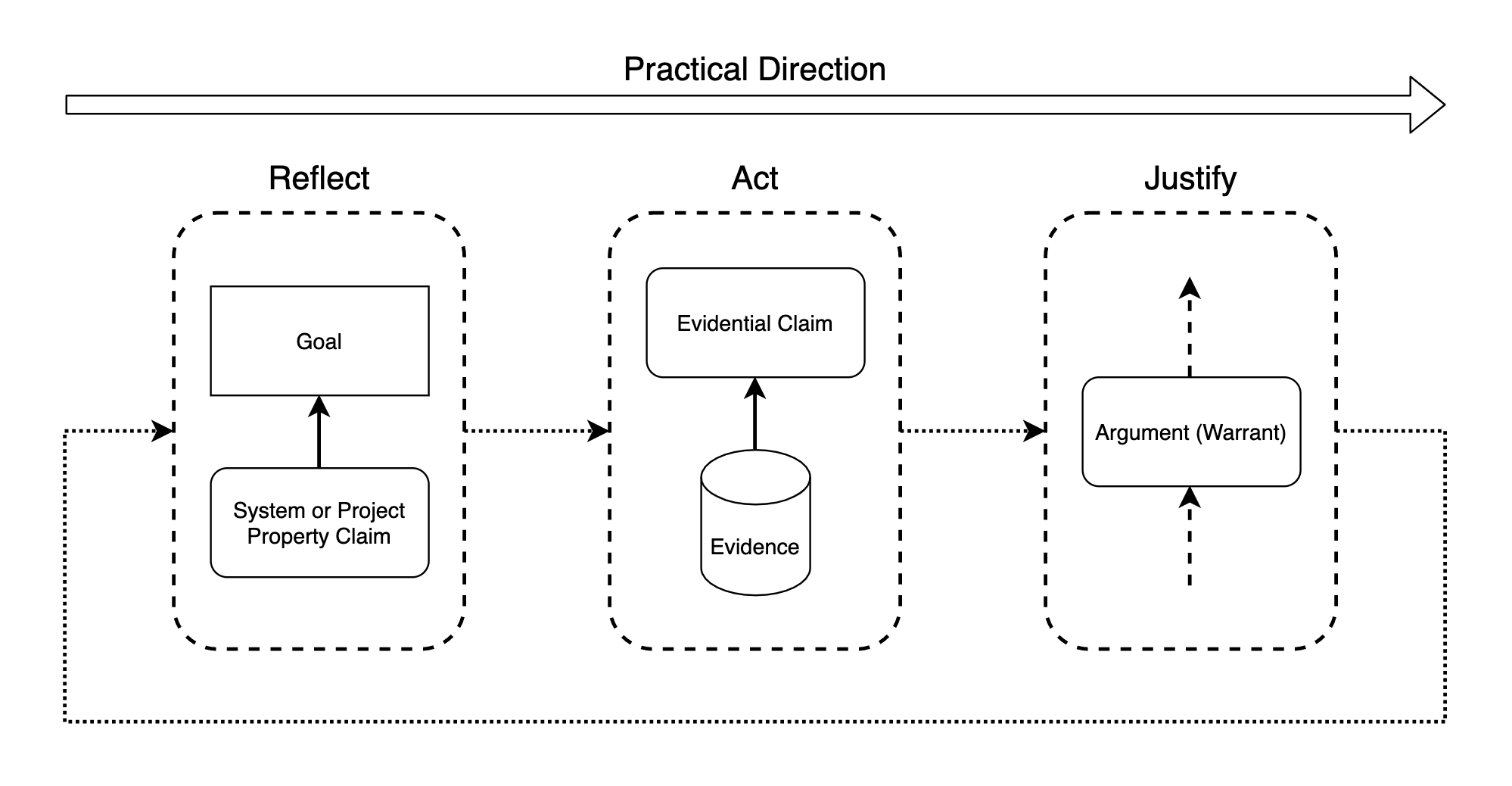}
\caption{A high-level schematic depicting the process of building an
ethical assurance case. The stages of reflect, act, and justify are
connected to the elements of an ethical assurance case.}
\end{figure}

This is captured in Figure 3, which presents a high-level schematic of
the over-arching process of reflect, act, justify, while also capturing
the various elements that are required in an ethical assurance case.

This schematic follows the \emph{practical direction} of project design,
model development, and system deployment. For instance, a project team
may begin in the \textsc{project planning} stage by identifying the
normative goal of sustainability, which requires \emph{inter alia} that
the practices behind the system's production and use be informed by
ongoing considerations of the potential for exposing affected people and
groups to harmful impacts. At the project design phase, operationalising
this goal will involve engaging in anticipatory deliberations about the
potential impacts of the project on the individuals and communities it
could affect. Such a process of impact assessment plays a vital role in
helping to set the direction of travel for the project (e.g., ensuring
that their system protects fundamental human rights and freedoms and
prioritises social justice) and in providing a shared vocabulary for
project team appraisal, stakeholder engagement, and public
communication. The action taken by the project team to realise the goal
of sustainability at the project planning stage, namely, the initial
impact assessment, provides evidence to justify the claim that, in the
project design phase, the goal of sustainability has been ascertained.

Sustainability is, of course, just one of many top-level normative goals
that may be identified and articulated by project teams as they reflect
on the salient ethical principles needed to assure the responsibility of
their AI system or ML model and the trustworthiness of the practices
behind building and using it. Key values associated with AI ethics also
help with the specification of the ethical principles that serve as the
top-level normative goals for the project---akin to the goal of safety
in a traditional assurance case. With these goals set, the project team
can then identify the necessary properties of the project or system that
must be established throughout the design, development, and deployment
of the system, and subsequently assured to justify how the relevant goal
has been obtained.

This \emph{anticipatory} process supports a responsible and ethical
approach to project governance, but also speeds up the process of
building an ethical assurance case. The decisions and actions that the
project team take at each stage of the project lifecycle will invariably
alter and refine their initial, anticipatory reflections, requiring a
cyclical and interactive process of reflection, action, and
justification. However, having a rough idea of what is required to
ensure that the fairness or the explainability of a system is
sufficiently established can support a phased approach to building an
assurance case, instead of leaving it to the final part of the project.

The following sections will follow this practical direction of building
an assurance case. To illustrate the role of each element within an
ethical assurance case, and to further motivate the value of the
`reflect, act, justify' process, we will use the running example of a
hypothetical project to design and develop a decision support tool to be
deployed in a healthcare setting.

The decision support tool uses a ML algorithm to triage (classify)
incoming patients on the basis of their observable symptoms and
physiological measurements (\(X\)), in order to determine their expected
risk of clinical deterioration (\(Y\)), and offer tailored guidance to the relevant
healthcare practitioner. For the purpose of this illustration we do not
need to worry about the specific details of \(X\) or \(Y\).\footnote{In formal terms, we can describe the task of a classifier as trying to determine (or, predict) the value of some unknown variable \(y_i\) \(\in\) \(Y\) based an an observed variable \(x_i\) \(\in\) \(X\). In the case of supervised learning, the ML algorithm is trained on a series of labelled data, taking the form \((x_1,y_1),...,(x_n,y_n)\), where each example is a pair \((x_i,y_i)\) of an instance \(x_i\)\hspace{0pt} and a label \(y_i\). The goal is to learn an optimal mapping function (given certain pre-specified constraints) from the domain of possible values for \(X\) to the range of values that the target variable \(Y\) can assume. This formulation of the classification task covers many concrete examples and algorithm types, at a high-level of abstraction (e.g., risk assessment, automated credit scoring, object identification). The general format of this case study is similar to many widely-used scoring systems, which need not rely on ML to function (e.g. \protect\hyperlink{ref-royalcollegeofphysicians2017}{Royal College of Physicians 2017}).}

Before the project team begin to design, develop, or deploy the
technology, they reflect upon which ethical goals they are trying to
achieve.

\hypertarget{top-level-normative-goal}{%
\subsubsection{Top-Level Normative
Goal}\label{top-level-normative-goal}}

As previously noted, an ethical assurance case begins with the
identification of \emph{top-level normative goals}, which are oriented
towards key ethical values (see
\protect\hyperlink{foundational-values}{§3.2}).

For example, our hypothetical project team are motivated (in part) by
the recognition that their tool could lead to an unequal and unfair
distribution of health outcomes among their target population, in virtue
of a) its discriminatory ability to classify patients (e.g.~based on
demographic, phenotypic, or physiological characteristics), and b) its
potential to influence the healthcare professional's diagnosis and
subsequent recommendation for treatment. However, by considering a range
of ethical values and principles, they also recognise that something
like `promoting greater health equity' may be insufficient on its own,
as good healthcare depends on the normative ideal of \emph{participatory
decision-making} between a healthcare professional and patient. In other
words, to avoid inappropriate levels of paternalism from healthcare
professionals, healthcare depends on a \emph{respect for patient
autonomy} and a \emph{need for informed consent} to proposed healthcare
interventions. Therefore, the team recognise at the outset that any
argument they proffer to justify a claim regarding greater health equity
must also incorporate considerations of how their system supports a
complementary principle (and goal) such as `explainability.'

\begin{figure}
\centering
\includegraphics{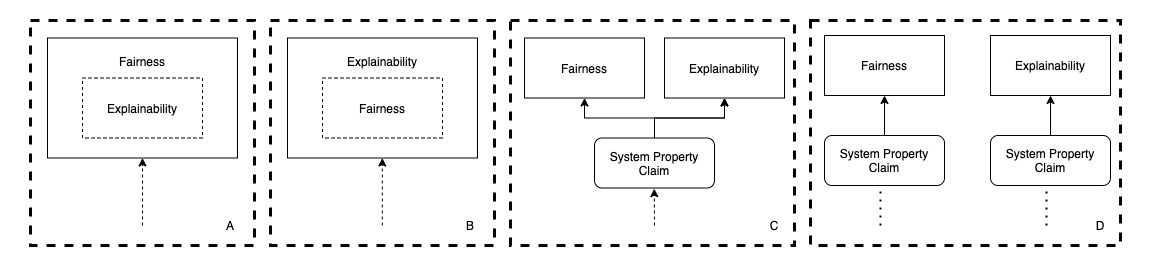}
\caption{(Non-exhaustive) options for presenting an argument that is
oriented towards multiple ethical goals. In this instance, the options
present a case where the goal of explainability is subsumed within a
higher-level goal of fairness (a), vice-versa (b), a case where both
goals are treated as jointly important and interlinked in virtue of the
lower-level claims they depend upon (c), and an option that builds two
separate assurance cases (d).}
\end{figure}

We make no specific claims about how these two complementary goals ought
to relate in any given assurance case, as they are merely illustrative
and the specification of ethical principles requires careful
consideration of context. However, Figure 4 shows how the two goals
could, in principle, relate to each other. The options are (a) fairness
could be the primary normative goal but depend upon the realisation of a
sub-goal of explainability, perhaps due to the importance of informed
consent; (b) vice-versa; (c) both goals are jointly important but
interlinked in virtue of the lower-level claims they depend upon; or (d)
the two goals are independent and require two separate assurance cases.
For simplicity, we will treat the goal of fairness as dependant on
explainability for the purpose of our illustrative example. Moreover, we
acknowledge that other values and principles may be involved
(e.g.~accountability), but choose to focus on fairness and
explainability for simplicity of exposition.

Starting with the identification of relevant ethical principles supports
the kind of anticipatory reflection that is emphasised by work in
responsible research and innovation
(\protect\hyperlink{ref-stilgoe2013}{Stilgoe, Owen, and Macnaghten
2013}). In ethical assurance, this reflective process is guided by the
scaffolding of the project lifecycle (see Figure 2). For example, at the
\textsc{project planning} stage, the project team may identify that they
need to consider how relevant social determinants of health could affect
the accuracy of the system for certain groups
(\protect\hyperlink{ref-lucyk2017}{Lucyk and McLaren 2017}), or whether
existing social biases that favour other sub-groups may lead to unfair
levels of access to the system. If the system is intended for use in an
area where a high percentage of patients are exposed to poor working
conditions, an important reflective question to ask would be, `Does the
model include relevant variables that can measure the relevant social
risk factors?' If the answer to this question is `no,' the system may
fail to accurately assess this sub-group. However, if a gap like this is
identified early enough, perhaps as a result of engaging relevant
experts or stakeholders during the \textsc{problem formulation} stage,
or during exploratory \textsc{data analysis}, then it may be sufficient
to alter plans for the \textsc{Data Extraction and Procurement} stage
that aim to improve the \emph{representativeness} of the dataset. If so,
the team could consequently select a relevant fairness optimisation
constraint that promotes greater health equity throughout the target
population, and which could be integrated during \textsc{(model)
development} (e.g., during pre-processing, training, or
post-processing), and then verified during \textsc{model testing \&
validation} and reported on during \textsc{model reporting}.
Alternatively, the process of identifying the top-level goal may lead to
a realisation that the function and purpose of the tool is poorly
understood, or that the potential benefits are outweighed by the
potential risks. If so, then the right decision may be to not proceed
with the project at all.\footnote{This is just a selection of
  considerations. We cannot hope to cover all other relevant topics
  here, such as the importance of ensuring that the fairness
  optimisation constraints are considered reasonable by the affected
  stakeholders.}

Therefore, by using the project lifecyle as a guide at the outset of a
project, the project team can reflect upon the possible decision points
and challenges that are likely to emerge throughout the project, and
which have a bearing on the selected goal.

In defining a top-level normative goal, it is not enough to simply state
that the goal is `fairness' or `explainability.' This is because a
top-level normative goal that simply stated, ``This decision support
tool is fair.'' would be \emph{insufficiently specified}, and would
simply give rise to the question, `What notion of fairness is being
employed?' Therefore, it is important to acknowledge how the
\emph{context} of the project provides the basis for the
\emph{specification} of the principles or goal (see
\protect\hyperlink{operationalising-normative-principles}{§3.3}). For
example, if the system in our hypothetical example were to be deployed
in a context of mental healthcare, where the target population lacked a
capacity for decisional autonomy, the principle of explainability would
be specified in a very different manner than if the system were to be
deployed in the case of cancer treatment among consenting adults.

In the case of our hypothetical example, we can assume that a first pass
at a normative goal would be something like the following:

\begin{quote}
The use of the decision support tool helps advance health equity.
\end{quote}

However, this is missing several vital pieces of information, and can be
made clearer by highlighting the relevant components of the goal. A
better version would be:

\begin{quote}
The use of the \textbf{\{decision support tool\}} by
\textbf{\{healthcare professionals in a formal healthcare setting\}} can
help advance \textbf{\{health equity\}}.
\end{quote}

Here, the brackets respectively highlight several important elements
that feed into the top-level normative goal:

\begin{itemize}
\tightlist
\item
  a description of the \textbf{\{technological system\}},
\item
  the \textbf{\{context\}} in which the system is being deployed, and
\item
  the \textbf{\{normative goal\}} that centres the assurance case
\end{itemize}

\begin{figure}
\centering
\includegraphics{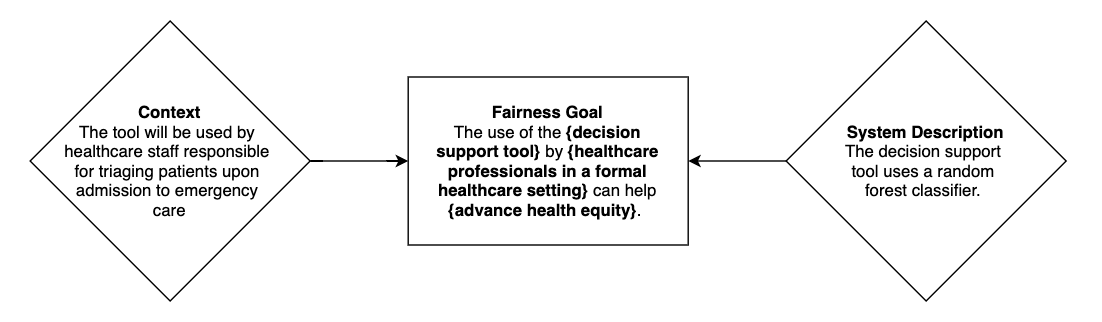}
\caption{An example of how the top-level normative goal can be further
contextualised and specified through supporting elements.}
\end{figure}

This allows further details for the components to be given in the linked
elements, as displayed in Figure 5. For instance, providing a short
description of the type of ML algorithm that is implemented within the
\{technological system\} or providing further specification to help
clarify the \{normative goal\} (e.g.~a definition of the notion of
`fairness' or `health equity' that is being employed). As the subsequent
components provide further detail that build on this initial goal,
keeping the formulation of the top-level goal concise is
advised.\footnote{This also connects with some possible, future
  directions for ethical assurance that we discuss in section
  \protect\hyperlink{conclusion-challenges-open-questions-and-next-steps}{6}.
  Specifically, the possibility of modularising ethical assurance in
  order to support the development of argument patterns or a model-based
  approach.}

\hypertarget{project-and-system-property-claims}{%
\subsubsection{Project and System Property
Claims}\label{project-and-system-property-claims}}

Once the top-level normative goal has been sufficiently specified, it is
then necessary to identify the actual properties of the project and
system that help operationalise the set of ethical principles that
define the goal of the project. This includes identifying the decisions
and actions taken throughout the project's lifecycle that ensure the
goal is actually achieved (e.g.~robust information governance processes
to protect sensitive healthcare information). Again, this stage can (and
should) proceed as an anticipatory process of reflection at the outset
of the project to help identify, assess, and respond to the possible
challenges that may arise throughout the project lifecycle.

In the case of our decision support tool, the project team may have
identified several (non-exhaustive) properties of their system or
project that are relevant to their goal, using the project lifecycle as
a guide. These properties can be formulated as statements about actions
or decisions that need to be taken during specific stages:

\begin{quote}
``During exploratory \textsc{data analysis} we must consider the
possibility that \emph{diagnostic access bias}\footnote{Diagnostic
  access bias arises when individuals differ in their geographic,
  temporal, and economic access to healthcare services, this variation
  may result in their exclusion from a study or dataset, differential
  access to diagnostic tests, or affect the accuracy of the diagnostic
  test itself. This can cause under- or over-estimation of the true
  prevalence of a disease, and lead to worse treatment for
  socioeconomically deprived individuals.} has affected the quality of
our training data.'' ``At the \textsc{model reporting} stage, it will be
important to ensure that information about the representativeness of our
dataset is recorded, while also remaining sensitive to the need to
maintain data privacy. Therefore, we will need to decide how granular to
make our data while remaining sensitive to any potential trade-offs with
accuracy metrics among patient sub-groups.'' ``The healthcare
professionals should be able to investigate and challenge the rationale
for a particular assessment outcome during \textsc{system use and
monitoring}, to ensure that professional judgement acts as a safeguard
against false positives and false negatives, while also supporting
explainability.''
\end{quote}

There are, of course, many more decisions that are relevant to the
top-level normative goal---our aim here is just to highlight some
illustrative examples. However, the process of using the project
lifecycle to support anticipatory reflection can help to uncover the
properties of the project or system, and identify possible actions that
could be undertaken.

For example, let's say the project team decide to consult a group of
clinical experts and social care workers with knowledge of the site from
which the data were generated to determine the likelihood that the risk
of \emph{diagnostic access bias} has been minimised, as well as any
other significant statistical, social, or cognitive biases. They could
then formulate the following claim about a property of their project:

\begin{quote}
``We consulted a panel of experts to independently assess our dataset
and ensure that the effect of bias has been minimised prior to model
development''
\end{quote}

This claim may be a \emph{necessary} component in developing a
convincing assurance case, but is insufficient on its own to justify the
top-level goal for at least two reasons.

First, additional claims will be required to \emph{jointly satisfy} the
top-level goal. Some of these claims may be subordinate to the parent
claim (e.g., claims about which types of bias were assessed). And, some
of these claims (as above) refer to aspects of the \emph{project's
management} (e.g.~choices about how the project was managed), rather
than to aspects of the system itself (e.g.~details about the user
interface of the decision support tool). Both sets (i.e.~project and
system claims) reflect important sources of procedural claims that aim
to legitimise the project's overall governance. This is why the project
lifecycle detailed in Figure 2 is such an important starting point for
the anticipatory and reflective processes that underpin ethical
assurance---it helps scaffold a thorough evaluation of both the system
itself and the processes that led to its design, development, and
deployment.

Second, the claim requires evidential backing, which we turn to next.

\hypertarget{evidential-claim-and-artefact}{%
\subsubsection{Evidential Claim and
Artefact}\label{evidential-claim-and-artefact}}

Once the project or system property claims, which \emph{jointly support}
the top-level normative goals, have been identified, it will be
necessary to link some of these claims to documented evidence, or
`evidential artefacts' (e.g., a report that details the outputs from the
\textsc{model testing or validation} stage). Developing the project and
system claims, and gathering the necessary supporting evidence, will, in
practice, be a simultaneous process---again, highlighting the iterative
and cyclical nature of the reflect, act, and justify process.

Whether the formulation of a system or project property claim requires
evidence will, of course, depend on the nature of the claim. As
Cartwright and Hardie (\protect\hyperlink{ref-cartwright2012}{2012, 53})
acknowledge,

\begin{quote}
``Some claims are self-evident or already well established. They do not
need to be backed up by anything further for you to be justified in
taking them to be true.''
\end{quote}

In these cases, there may be no need to document an `evidential
artefact.' Whereas, in other cases an evidential artefact will need to
be referenced and an `evidential claim' will need to be established. An
evidential claim can be treated as a proposition or description of the
relevant evidence.

The intended audience of the assurance case itself will play a part in
determining whether evidence is required, and how the evidential claim
is formulated. Here, trust may further play a role in whether a claim is
accepted without evidence. For instance, if the party responsible for
developing an assurance case is trusted by the party reviewing it, they
may be willing to accept a propositional claim as evidence, rather than
demanding further documented evidence (e.g.~an external auditor versus
an internal red team).

\begin{figure}
\centering
\includegraphics{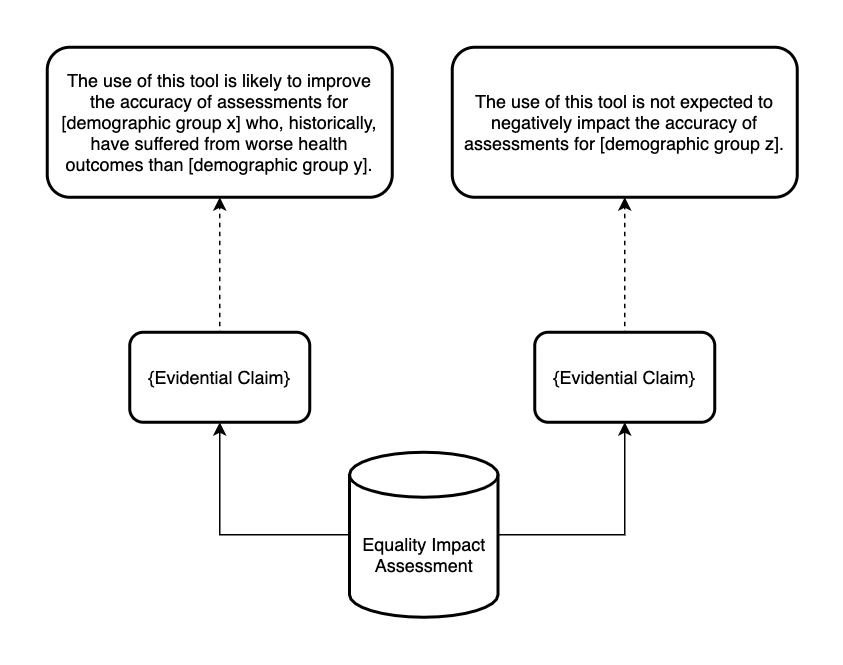}
\caption{A portion of an assurance case showing how two claims about a
project can be supported by the same evidential artefact.}
\end{figure}

The evidence may also substantiate multiple claims if the system
property claim and argument are wide in scope. As such, there may be a
many-to-one relationship between evidential claims and an artefact. For
example, consider the two system/project claims about our hypothetical
project depicted in Figure 6. As supporting evidence for these two
claims, the assurance case may refer to the findings of an equality
impact assessment undertaken at the outset of the project, which in this
case may serve as a single evidential artefact that helps justify the
two claims (with appropriate reference to specific sections of the
assessment). This is important, as it means that certain pieces of
documentation, which may be generated during typical project activities,
such as algorithmic impact assessments
(\protect\hyperlink{ref-reisman2018}{Reisman et al. 2018};
\protect\hyperlink{ref-leslie2019}{Leslie 2019}), transparency
statements (\protect\hyperlink{ref-collins2015a}{Collins et al. 2015}),
or datasheets (\protect\hyperlink{ref-gebru2018}{Gebru et al. 2018})
could help ground several of the claims beign made. In turn, regulators
could incentivise or mandate certain activities that are important
sources of evidence for ethical assurance cases.

However, in many cases, establishing a complete justificatory link
between a system/project claim and an evidential claim requires one
final step.

\hypertarget{argument-warrant}{%
\subsubsection{Argument (Warrant)}\label{argument-warrant}}

The final step in the practical process of building an ethical assurance
case is more nuanced than the previous steps but is absolutely crucial
for making key assumptions explicit, and ensuring that the standpoint on
which the argument depends is clearly demonstrated.

Consider the following two claims:

\begin{quote}
\textbf{System/Project Property Claim}: the \textbf{\{ML model\}}
produces fair outputs because it has been trained with
\textbf{\{fairness optimisation constraint x\}}. \textbf{Evidential
Claim}: the results of the \textbf{\{fairness optimisation process\}}
were \textbf{\{\ldots\}}.
\end{quote}

The role of the evidential claim here is intended to support the
system/project property claim. However, whether an individual is
justified in assenting to a belief in the system claim depends on a
missing assumption between the two claims---whether the fairness
optimisation process is reliable or appropriate.

The philosopher Stephen Toulmin
(\protect\hyperlink{ref-toulmin2003}{2003}), whose work heavily
influenced the development of argument-based assurance, referred to this
link as the `warrant.' Simply put, the warrant is a step in an argument
that links some evidence to a particular propositional claim, with the
possible addition of a qualifier. This is indicated in Figure 7 with a
toy example.

\begin{figure}
\centering
\includegraphics{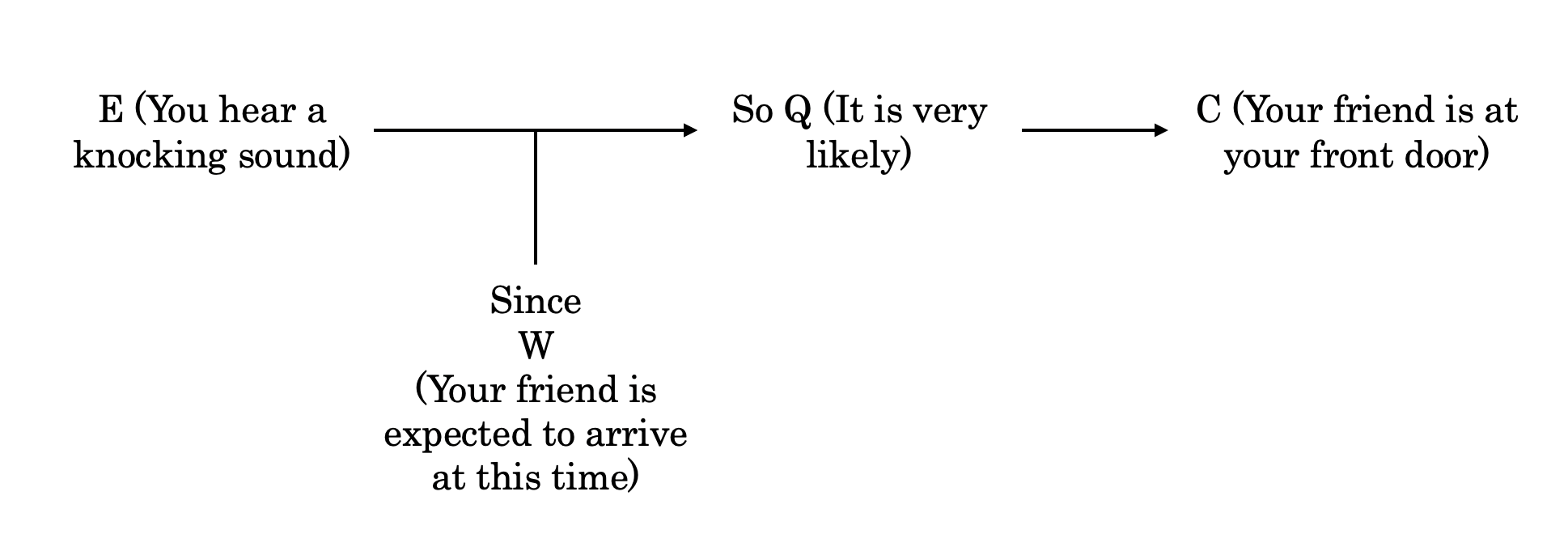}
\caption{Diagram showing the relationship between several propositions:
evidence (E), warrant (W), and a qualified (Q) claim (C).}
\end{figure}

In the above example, the inference from the evidence (You hear a
knocking sound) to the qualified claim (It is very likely your friend is
at your door) is only justified when we also provide additional warrant
(Your friend is expected to arrive at this time). Identifying the
warrant that supports the inferential step from evidence to a claim can
be one of the most challenging and complex steps in developing an
assurance case, as it requires the project team to make explicit many of
their assumptions.\footnote{It also involves what epistemologists refer
  to as the \emph{transmission of justification across inference}, which
  is a process where the justification for one belief (p) derives its
  justification from the justification that one has for a secondary
  belief (q) (\protect\hyperlink{ref-moretti2013}{Moretti and Piazza
  2013}).} For example, let's assume our hypothetical project team
decide to adjust their classifier during post-processing so that it is
uncorrelated with a protected attribute, and subsequently provide
evidence about the model's performance. To establish a link between the
evidential claim about the model's evaluation and a system claim
relating to, say, group fairness, the project team will also need to add
something like the following:

\begin{quote}
``The use of {[}fairness optimisation constraint x{]} is legitimate in
this context as it was selected by a representative group of
stakeholders and experts, during a process of deliberative dialogue, as
the most appropriate technique for operationalising and embedding our
notion of fairness within the project.''
\end{quote}

In practice, the project team may have to construct the full argument
once the surrounding components (i.e.~`system property claim' and
`evidential claim') have all been selected, and the inferential links
between the set of claims is apparent. Although there is a concern that
this may lead to the construction of \emph{post hoc} arguments, it may
also be necessary to fully evaluate whether the selected evidence is
sufficient to support the argument (see
\protect\hyperlink{determining-and-evaluating-evidence}{§4.4.1} on
determining and evaluating evidence). For instance, the project team may
realise that a link between two claims rests on a faulty assumption, and
that further work will be required. Again, the sooner that an
anticipatory and ongoing process of reflection, action, and
justification begins, the more likely such gaps will be identified and
addressed.

Now that we have seen the practical direction for building an assurance
case, we are in a better position to understand the complementary
justificatory direction that links the elements together.

\hypertarget{the-justificatory-direction}{%
\subsection{The Justificatory
Direction}\label{the-justificatory-direction}}

\begin{figure}
\centering
\includegraphics{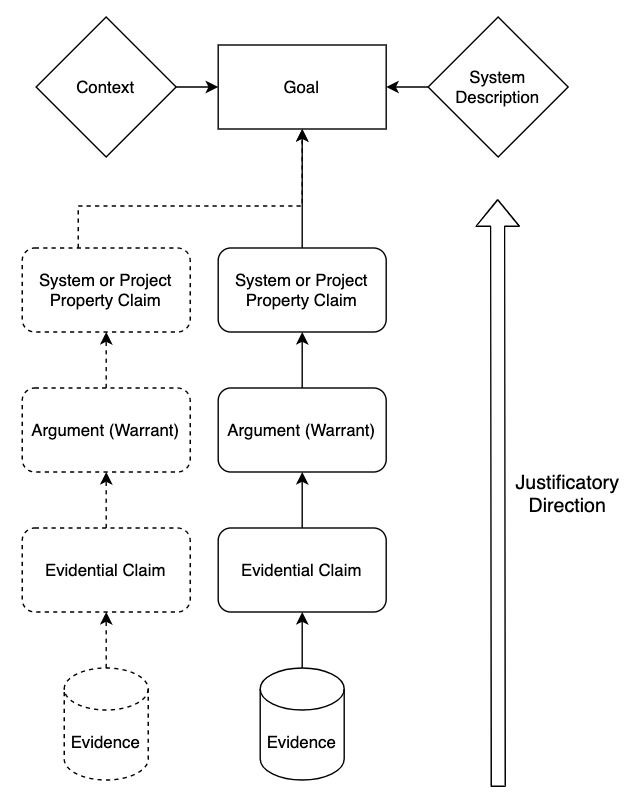}
\caption{The general structure of an ethical assurance case, indicating
the various elements and their relationship to each other.}
\end{figure}

Figure 8 offers a representation of the logical structure that links
each of the elements together. The directionality of the arrows
represents the fact that the lower-level components provide
\emph{inferential support} for the higher-level claims (e.g.~an argument
that warrants a system property claim; an evidential claim that raises
the likelihood of an argument being valid). This is why the arrows lead
upwards, or are oriented towards the top-level normative
goal.\footnote{Those readers who are familiar with informal logic and
  argumentation theory will recognise that this structure is also
  heavily influenced by the work of Stephen Toulmin
  (\protect\hyperlink{ref-toulmin2003}{2003}), whose research into the
  structure of arguments has been highly influential in the development
  of ABA.} As can be seen, the structure that we propose is closely
modelled on \emph{goal-oriented} tools such as GSN or CAE.

In addition to this goal-oriented focus, ethical assurance also has a
systems-level focus. Like traditional safety cases, the complexity of an
ethical assurance case will depend on the complexity of the system
itself (and the social context in which it is deployed), and the level
of detail required will be guided to some degree by a principle of
proportionality. Ethical assurance departs from tools like GSN, however,
in that it aims for representational simplicity rather than
completeness. As such, it has a lot less \emph{representational
expressivity}. This is a purposeful design consideration to improve
accessibility for the purpose of supporting inclusive, participatory
methods.

\begin{figure}
\centering
\includegraphics{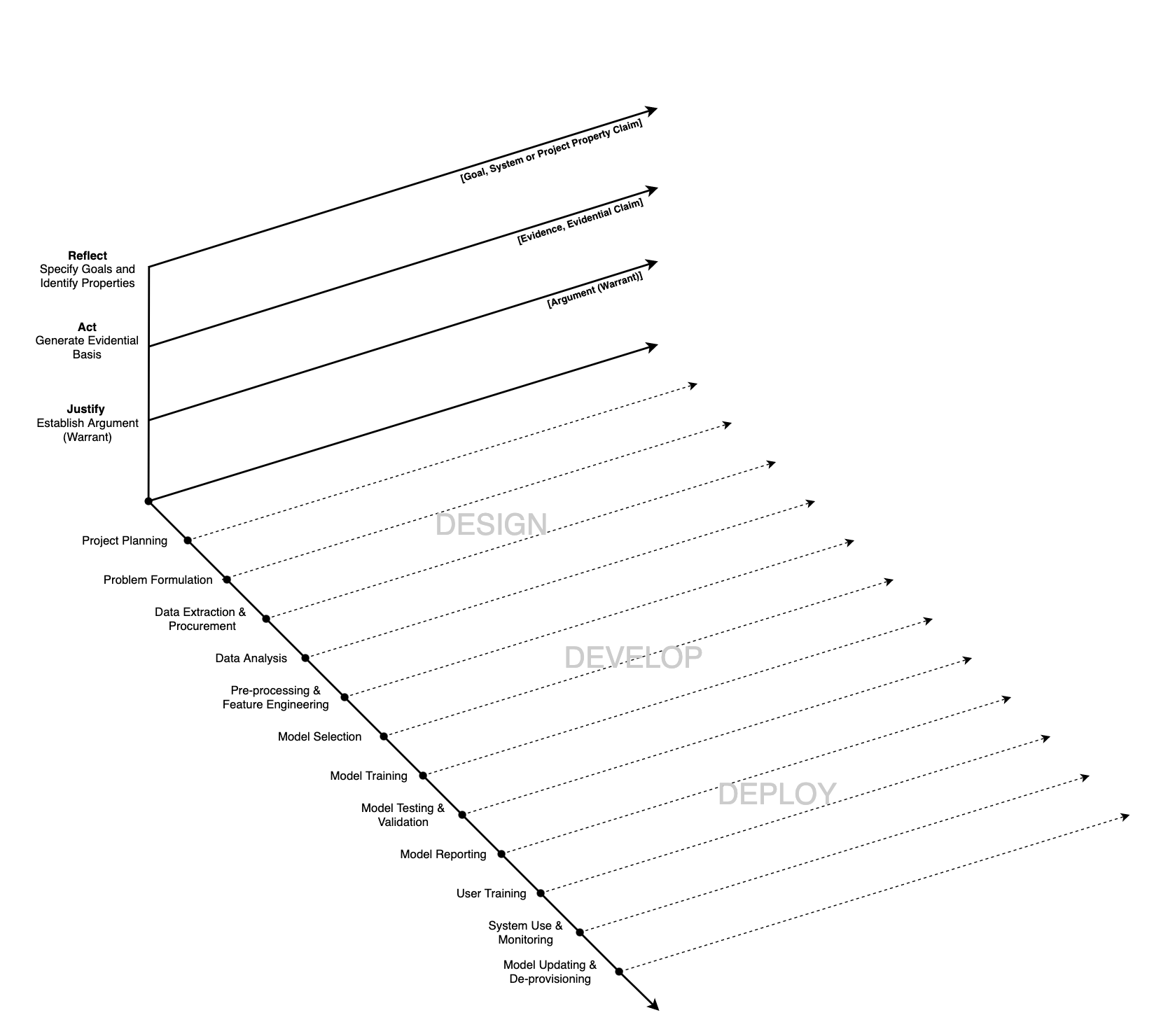}
\caption{A unifying schematic showing the iterative `reflect, act,
justify' process connected with the elements of ethical assurance case,
and grounded in the project lifecycle.}
\end{figure}

Figure 9 ties all these elements, structure, and processes of an ethical
assurance case together into a single schematic, further demonstrating
the value of using the project lifecycle model as a tool for scaffolding
an anticipatory process of reflection, action, and justification.

\hypertarget{building-and-using-an-ethical-assurance-case}{%
\subsection{Building and Using an Ethical Assurance
Case}\label{building-and-using-an-ethical-assurance-case}}

In closing this section, we expand on three additional topics regarding
the development of an ethical assurance case and also discuss the role
an ethical assurance case plays once it has been built.

\hypertarget{determining-and-evaluating-evidence}{%
\subsubsection{Determining and Evaluating
Evidence}\label{determining-and-evaluating-evidence}}

As we have demonstrated, the process of developing an ethical assurance
case relies on the collection and use of evidence. But how does one go
about \emph{evaluating} the probative value of evidence in the context
of ethical assurance? And what counts as evidence in the first place?
Drawing upon concepts from argumentation theory and jurisprudence can
help answer these questions, especially given our context of
\emph{argument-based} assurance.

Recall that what is being developed here is a \emph{structured} case
that can be used to support a process of \emph{reason-giving} (on the
part of the project team). This act of giving reasons implies a
recipient and an active process of dialogue---even if it is asynchronous
and mediated through a formal process, rather than a verbal one. Who the
recipient of an assurance case will be will obviously depend. However,
in spite of this, we can treat the exchange of reasons, structured as an
assurance case, as similar to the determination, use, and interrogation
of evidence in a legal setting (e.g.~arguing a case to a judge or jury,
which is critiqued by another party).

Three principles guide fact finding in a legal case: \emph{relevance},
\emph{materiality}, and \emph{admissibility} (see
\protect\hyperlink{ref-ho2015}{Ho 2015} for a more detailed exposition).
`Relevance' is a relational concept that holds between two propositions.
In the present context, this is the relevance of the evidence in
establishing (qualified) warrant for the respective claim. `Materiality'
refers to whether or not a fact is receivable by a court, which in turn
depends on additional legal facts. For instance, a fact that is not in
dispute may be \emph{relevant} in the ordinary sense of the term, but
not material to a court's decision (e.g.~a fact about a breach of
contract that is accepted by both parties). The materiality of evidence
is significant because it reduces the need to consider (relevant)
evidence that will not affect the outcome of the (assurance) case.
`Admissibility' is based on the rule of law and covers conditions for
exclusions that go beyond relevance or materiality. For instance, a fact
may be both relevant and material but still be inadmissible due to
additional legal rules (e.g.~hearsay is not admissible even if it is
relevant and material).

Similar to the question of what constitutes relevant, material, and
admissible evidence in a court, we can ask `under what conditions can an
evidential claim or documented evidence serve as grounds for
\emph{reasonable inference} in an assurance case?' Here, we are aiming
to assess the \emph{probative value} of a particular evidential artefact
(or associated claim) within an assurance case. For example, anecdotal
evidence may be accepted in everyday conversation but would not be
admitted as evidence in court.

Furthermore, when evaluating the probative value of evidence within an
assurance case, we must do so with an eye to the whole case, not just
the isolated value of a particular claim or artefact. This requires
consideration of:

\begin{enumerate}
\def\labelenumi{\arabic{enumi}.}
\tightlist
\item
  The probative value of each evidential artefact/claim in relation to a
  specific project/system claim
\item
  The sufficiency of either an individual evidential artefact/claim or
  set of artefacts/claims in relation to a specific project/system claim
\item
  The sufficiency of the overall assurance case conditional on the
  top-level normative goal\footnote{The sufficiency of the overall
    assurance case will, of course, depend on 1 and 2.}
\end{enumerate}

There is a further sense in which the evaluation of evidence is
relational. Whether a propositional claim or documented artefact counts
as evidence depends on whether it is judged by the \emph{recipient} as
supporting a process of reasonable inference. That is, in a process of
\emph{dialogical reasoning}, does the evidence serve as a reason in
support of the conclusion that a proposition (i.e.~the claim) is true or
false, probable or improbable, for the parties involved? Here, it is
important that the top-level normative goal is grounded in a value or
principle (e.g.~human dignity) that is accepted as a shared and public
grounds for communication. Hence, the need to base an argument on the
shared acceptance of reasonable ethical values and principles
(\protect\hyperlink{ref-leslie2019}{Leslie 2019}).

There is, understandably, no general answer that can be offered here
when it comes to the determination and evaluation of specific evidential
artefacts and claims. However, raising this question nevertheless
clarifies a previously mentioned function of ethical assurance that is
sometimes neglected in ML/AI assurance: the need to support active
enquiry.

\hypertarget{phased-assurance-and-active-enquiry}{%
\subsubsection{Phased Assurance and Active
Enquiry}\label{phased-assurance-and-active-enquiry}}

It is already well recognised that the method of ABA goes beyond the
mere documentation of development processes to support compliance and
regulatory obligations. Instead, the development of an assurance case
can, among other things, assist anticipatory \textbf{reflection} and
\textbf{deliberation} (as seen above), and also help build and establish
trust through \emph{transparent communication}. As an ethical standard
for algorithmic accountability, however, the goal of transparency has
been criticised on a number of grounds. For instance, some researchers
have pointed to legal and financial barriers to achieving transparency,
resulting in the need to develop and employ techniques from areas such
as computational journalism to get around these barriers and assess the
use of algorithmic systems in high-stake domains (e.g.~advertising,
criminal justice) (\protect\hyperlink{ref-diakopoulos2014}{Diakopoulos
2014}). Unfortunately, these techniques can be limited in significant
ways and transparency is not equivalent to governance or control
(\protect\hyperlink{ref-diakopoulos2015}{Diakopoulos 2015}). Other
researchers have questioned the ideal of transparency more generally,
arguing that it can be used as a means to intentionally occlude, as with
the case of a company that provides mountains of evidence to be sifted
through at the last minute prior to a legal case
(\protect\hyperlink{ref-ananny2018}{Ananny and Crawford 2018}). To help
address these concerns about transparency, we can turn to the work of
the moral philosopher Onora O'Neill.

In her BBC Reith Lectures, \emph{A Question of Trust}, O'Neill
(\protect\hyperlink{ref-oneill2002}{2002, 77--78}) argues that,

\begin{quote}
``if we want a society in which placing trust is feasible we need to
look for ways in which we can \emph{actively check} one another's
claims.''
\end{quote}

She also acknowledges,

\begin{quote}
``active checking of information is pretty hard for many of us.
Unqualified trust is then understandably rather scarce.''
\end{quote}

This notion of `active checking' is important. In the context of ethical
assurance we can think of it as a form of dialogical communication that
we label `active enquiry.' That is, the development of an ethical
assurance case, ought not be approached as a monological exercise, in
which a project team produce an argument that is simply presented as
fact to relevant stakeholders. Rather, each argument is
\emph{necessarily defeasible}, because of both its inferential structure
and the possibility that the acceptability or weighting of the normative
goal may be legitimately challenged by certain stakeholders. This
\emph{contestability} of an assurance case means it should be seen as a
living document that can form the basis for an ongoing conversation
about the acceptability of sociotechnical systems in key areas of
society.

This connects to the second point. The iterative, ongoing, and situated
nature of the design, development, and deployment of data-driven
technologies requires a \emph{phased approach} to ABA. Kelly
(\protect\hyperlink{ref-kelly1998}{1998}) refers to this as the
presentation of an ``evolving safety argument,'' outlining three stages:
a preliminary stage, interim stage, and operational stage.

In the first stage, only an outline of the argument is produced, showing
the principal objectives, and anticipated evidence---similar to our
presentation of anticipatory reflection in
\protect\hyperlink{ethical-assurance}{§4}. In the second stage, the
argument is developed to reflect the increased knowledge of the design
and specification of the system---similar to the embedded and continuous
reflection throughout the project lifecycle (see Figure 2). Finally, the
case evolves to reflect evidence that concerns how the system is tested
and implemented. This general strategy is embedded in the Goal
Structuring Notation (GSN) approach, but its influence should also be
clear in the reflect, act, and justify approach to ethical assurance
(see Figure 3). However, we must take Kelly's evolving safety argument
approach one step further, and join the two ends together, in order to
capture the cyclical and socially situated nature of ethical assurance.
In doing so, the evolving nature of phased assurance aligns with the
idea that moral deliberation and public ethical reasoning is best
understood as a process of \emph{reflective equilibrium}
(\protect\hyperlink{ref-rawls1999}{Rawls 1999}). As social norms and
practices evolve, so too do the ethical demands and expectations on
sociotechnical innovation. Exercising responsibility and maintaining
trust, therefore, requires an inclusive, deliberative approach that
remains responsive and open to new moral horizons---hence the spiral in
Figure 2.

The third and final point that emerges from O'Neill's analysis is
captured by the recognition that active enquiry is often limited by the
aforementioned epistemic barriers such as intellectual property rights
that restrict access to information, or limited technical literacy or
capacity that inhibits an individual's understanding of complex
technical systems. Moreover, as organisations start exploring automated
ways of testing, validating, and documenting ML development---such as
Google's automation of \textsc{model reporting} activities
(\protect\hyperlink{ref-fang2020}{Fang and Miao 2020})---there is a very
real risk that developers become more and more epistemically detached
from the project lifecyle. This could, in turn, reduce opportunities for
vital ethical reflection, such that assurance becomes an automated
process that churns out floods of documentation---this would represent a
form of ``unintelligent accountability,'' to appropriate a term from
O'Neill (\protect\hyperlink{ref-oneill2002}{2002}). Returning to the
raised concern about transparency, this would place a disproportionate
burden on individual users or groups of stakeholders to ``seek out
information about a system, to interpret that information, and determine
its significance'' (\protect\hyperlink{ref-ananny2018}{2018, 979}). To
counteract this possible trend, we need to continue to consider which
practical mechanisms best support ethical assurance in general and
processes of active enquiry and phased assurance more specifically.
There are many open questions about how this can be achieved (see
\protect\hyperlink{conclusion-challenges-open-questions-and-next-steps}{§6}.
In the next section, we suggest one mechanism of direct relevance to
ethical assurance.

\hypertarget{ethical-argument-patterns}{%
\subsubsection{(Ethical) Argument
Patterns}\label{ethical-argument-patterns}}

In \protect\hyperlink{argument-patterns}{§2} we introduced the idea of
argument patterns---\emph{reusable templates} for assurance claims,
which address the types of claims, evidence, warrant and additional
contextual information that must be covered to justify a claim
pertaining to a particular normative goal. Is it possible to develop
ethical argument patterns? If so, what role would they play?

Let's start with the first question. How could we develop ethical
argument patterns? Although patterns could be proposed in a
prescriptive, top-down manner, in the case of ethical assurance they are
likely to be have greater normative force and legitimacy if they arise
bottom-up from actual patterns that are identifiable and generalisable
from existing assurance cases. For instance, if several ethical
assurance cases, which were concerned with the goal of `protecting human
dignity' when deploying automated decision-making systems in criminal
justice, all converged on a shared argument structure, this would be
reasonable grounds for inferring that the structure represented a
reliable argumentation pattern. As such, the pattern could be used as a
starting point for subsequent assurance cases, and some of the elements
could perhaps take on the role of so-called ``default reasons''
(\protect\hyperlink{ref-horty2014}{Horty 2014}).

An alternative means by which patterns could be developed is through
stakeholder engagement activities. For instance, activities including
deliberative dialogue (\protect\hyperlink{ref-andersson2013}{Andersson
et al. 2013}), in which stakeholders consider relevant information from
multiple points of view, enables the exploration and discussion of
topics or issues, without assuming the prior existence of consensus or
agreement. Instead, the process is designed to \emph{build consensus}
such that any judgements on the topics or issues under discussion are
arrived at through careful consideration and a greater understanding and
awareness of possible tensions. These types of engagement activities
could support the development of argument patterns, by exploring issues
such as socially desirable ethical goals, acceptable means of
specification, convincing and persuasive arguments, and accessible forms
of evidence.

Working with stakeholder groups in this way would also help expand the
role of both ethical assurance and ethical argument patterns. Firstly,
deliberative dialogue is not simply a method of opinion polling or
aggregation (\protect\hyperlink{ref-dryzek2003}{Dryzek and List 2003}).
More than this, the focus on dialogue and consensus building helps
orient the activities towards a more dialogical mode of education, which
could help improve digital literacy for the participants. And, second,
by engaging stakeholders in the assurance process, they are more likely
to have trust in the systems being deployed, as they will either
recognise their own values in the assurance case/patterns, or at least
have a greater understanding of the tensions and trade-offs that have
been reflected upon throughout a project's lifecycle.

The three topics we have explored---determining and evaluating evidence,
phased assurance and active enquiry, and ethical argument
patterns---help demonstrate the potential value of ethical assurance. We
now turn to consider some possible challenges, open questions, and next
steps.

\hypertarget{conclusion-challenges-open-questions-and-next-steps}{%
\section{Conclusion: Challenges, Open Questions, and Next
Steps}\label{conclusion-challenges-open-questions-and-next-steps}}

The proposals we have offered in this article are still very much in
their nascency, but we believe that ethical assurance has a lot of
promise for supporting and complementing ongoing attempts to ensure that
data-driven technologies work to promote individual and social
well-being. To further strengthen the support for this claim, in closing
this paper we anticipate some potential challenges for ethical assurance
and offer some responses. We also note the need for further research and
suggest next steps that we hope others will choose to contibute to, in
order to help turn this proposal into an active programme of research
and development.

\hypertarget{challenges}{%
\subsection{Challenges}\label{challenges}}

\hypertarget{ethical-assurance-could-be-misused}{%
\subsubsection{Ethical assurance could be
misused}\label{ethical-assurance-could-be-misused}}

This worry is reflected in claims that AI ethics supports cases of
``ethics-washing'' (\protect\hyperlink{ref-hao2019}{Hao 2019}) or as PR
coverage for ethically problematic institutional practices. The worry
can, of course, be extended to ethical assurance (i.e., ethical
assurance could be used as a mean for legitimising or covering up
unethical projects).

An immediate response is to simply note that this is a risk for nearly
all practical mechanisms that operate in this space, as tools can be
used for good and for bad. However, this response is unsatisfying. A
better response would be to note that the worry is itself mitigated by
the way in which ethical assurance has been designed.

By exposing the argumentation structure to open critique and active
enquiry, an ethical assurance case is more likely to expose an
unconvincing or incomplete argument. In turn, there is a potential for
improving the argument, or using available legal mechanisms to hold the
organisation accountable---an improvement on the limited notions of
transparency noted above. Moreover, the reflective and anticipatory
approach to the design, development, and deployment of algorithmic
systems that is enjoined by ethical assurance prevents the
superficiality of ``ethics washing'' practices in virtue of the
requirement that documented bridges be built between actions and
justifications across the entire ML lifecycle.

\hypertarget{ethical-assurance-is-too-demanding}{%
\subsubsection{Ethical assurance is too
demanding}\label{ethical-assurance-is-too-demanding}}

No one said ethics was easy. Doing the right thing can be challenging
and time-consuming, and in some cases the costs of failing to consider
ethics can be greater than the costs of doing the right thing in the
first place. Nevertheless, this is a valid concern when we consider the
limited capacity, available resources, and skills that often inhibit
smaller businesses and public sector organisations. Therefore, the
challenge here is to address how smaller organisations will manage the
increased demand that ethical assurance places upon them, alongside
existing regulatory requirements.

First of all, ethical assurance should not be misconstrued as a form of
compliance. It is not a separate requirement akin to a data protection
impact assessment, but rather a scaffold to support and supplement these
pre-existing requirements while promoting virtues such as transparent
communication and public or stakeholder accountability. While an
\emph{ethical assurance case} may take time to produce, the overarching
process has been designed to complement and extend existing regulatory
requirements and emerging best practices of design, development, and
deployment, rather than to be an additional burden to be completed at
the end of a project. This is why the project lifecycle is important as
an anticipatory guide.

However, it is still important to keep in mind a principle of
proportionality. Some projects are exposed to greater risk due to the
context in which they are developed (e.g., healthcare). Where a project
has very low risk, it may not be necessary to develop an extensive
ethical assurance case, but instead just use the methodology to guide a
process of reflection and document this accordingly. Here, argument
patterns may serve a further role by helping developers assess and
evaluate possible risks or benefits during the \textsc{project design}
stage.

In addition, ethical assurance will be made less demanding by improving
the skills and capacities of regulators and auditors, perhaps exploring
ethical standards and certification schemes that complement ethical
assurance (e.g., as forms of trusted evidence) and make the process more
efficient. Ultimately, ethical assurance is a framework that can be made
increasingly easy to employ as additional supporting mechanisms emerge.

\hypertarget{ethical-assurance-cases-are-defeasible}{%
\subsubsection{Ethical assurance cases are
defeasible}\label{ethical-assurance-cases-are-defeasible}}

This challenge focuses on the inferential support that the evidence
provides for both the property/system claims and the top-level normative
goal. As such, the overall argument will be defeasible, remaining open
in principle to revision, based on objections or competing forms of
evidence. This may seem, prima facie, like a problem for the
methodology. However, on deeper reflection it becomes clear that this is
a feature (and not a bug), which supports the function and purpose of
assurance.

As we discussed in §\href{from-safety-assurance-to-ethical-assurance}{3}
and §\protect\hyperlink{phased-assurance-and-active-enquiry}{4}, ethical
assurance is designed to support a process of reflection, action, and
justification throughout a project's lifecycle, and also enable the
active enquiry of stakeholders both prior to and after a system is
implemented within a particular context. For many technologies, their
social impact will not be made apparent until the time of deployment. A
method of assurance that failed to acknowledge these uncertainties would
fail to exercise an appropriate level of epistemic humility. In
contrast, the defeasibility of an ethical argument can a) help motivate
the need for an inclusive and participatory approach to design,
development, and deployment, b) encourage developers to adopt a phased
and modular approach to building an assurance case, and c) ensure that
possible visions for ethically- and socially-desirable futures remain
open to consideration.

\hypertarget{open-questions}{%
\subsection{Open Questions}\label{open-questions}}

\hypertarget{how-should-developers-deal-with-sensitive-or-probabilistic-evidence}{%
\subsubsection{How should developers deal with sensitive or
probabilistic
evidence?}\label{how-should-developers-deal-with-sensitive-or-probabilistic-evidence}}

This is in fact two questions, both of which pertain to the use of
evidence: How should sensitive evidence that can not be made public be
incorporated into an assurance case? How should probabilistic evidence
be used and evaluated?

In the case of sensitive evidence, it will be important to develop the
capacity of independent auditors, to ensure appropriate levels of
transparency. Likewise, it will be important for producers of ethical
assurance cases to develop layered or tiered methods of presentating
their cases, so that sensitive information can differentially be made
available to relevant parties and the that needs of non-technical
stakeholders can be accommodated by a plain language and easily
understandable layer of presentation
(\protect\hyperlink{ref-ico2020}{ICO and Institute 2020}) (also see
\protect\hyperlink{systems-and-standards}{§5.3.2}).

However, for both of these questions, we must also encourage the
development and adoption of norms and best practices within and across
domains and industries. For instance, in the case of an autonomous
vehicle, current efforts to deal with probabilistic evidence to support
safety claims (e.g.~this vehicle has not caused an injury in \(x\)
number of miles or journeys) are based on the notion of reducing risk to
levels that are as low as reasonably practicable (known as the ALARP
principle) (\protect\hyperlink{ref-lawcommission2020}{Commission 2020}).
Industry norms play a role in determining what is ``reasonably
practicable,'' but it is not presently clear whether or how this
risk-based approach will transfer to the case of ethical assurance.

\hypertarget{how-does-ethical-assurance-work-for-projects-that-distribute-responsibilities-across-teams-and-organisations}{%
\subsubsection{How does ethical assurance work for projects that
distribute responsibilities across teams and
organisations?}\label{how-does-ethical-assurance-work-for-projects-that-distribute-responsibilities-across-teams-and-organisations}}

The stages of design, development, and deployment for a complex ML-based
system, and the tasks within these stages, may be carried out by
different teams, across multiple organisations. For instance, the
increasing availability of so-called ``model libraries''---repositories
for pre-trained models---means that the procurement of components may go
beyond that of just the data or other services necessary for deploying a
system. As such, the evidence required to fulfil an ethical assurance
case may require the coordination of multiple actors.

Research into modular assurance cases remains an open question even in
the comparatively well-established safety case literature
(\protect\hyperlink{ref-habli2021}{Habli, Alexander, and Hawkins 2021}).
However, there is good reason to think that, in conjunction with a
phased approach to development, a modular form of ethical assurance
cases could support distributed project management. In addition,
assurance contracts, which hold parties legally accountable for the
claims made within their ``module'' (e.g., that they are not falsifying
evidence) could also be explored to support more complex cased and
maintain trust.

\hypertarget{next-steps}{%
\subsection{Next Steps}\label{next-steps}}

\hypertarget{evaluating-efficacy-of-ethical-assurance}{%
\subsubsection{Evaluating Efficacy of Ethical
Assurance}\label{evaluating-efficacy-of-ethical-assurance}}

Ethical assurance is a type of argument-based assurance, and, therefore,
has many of the same benefits that were outlined in
\href{the-structure-elements-and-properties-of-an-assurance-case}{§2.3}.
However, the extent to which these benefits accrue to projects that
employ assurance cases is a matter that remains inconclusive. For
instance, to what extent do safety cases contribute to the
safety-related outcomes of a project? Those who question the efficacy of
safety cases can point to well-known failures, such as the loss of the
RAF Nimrod MR2 Aircraft XV230 in Afghanistan in 2006, in support of
their skepticism. In an independant review of this incident, the safety
case that was drawn up for the aircraft by BAE systems was described as
``a lamentable job from start to finish. It was riddled with errors. It
missed the key dangers. Its production is a story of incompetence,
complacency, and cynicism.''
(\protect\hyperlink{ref-haddon-cave2009}{Haddon-Cave et al. 2009, 10})
Findings such as these do not appear to paint a positive picture for the
efficiacy of assurance cases. However, it would be premature to write
ABA off altogether.

It is clear from the Nimrod review that part of the failure stems from
organisational malaise among the producer of the safety case. As the
review notes, ``the task of drawing up the Safety Case became
essentially a paperwork and `tick-box' exercise.''
(\protect\hyperlink{ref-haddon-cave2009}{Haddon-Cave et al. 2009, 10})
In earlier sections we have been careful to acknowledge that the
production of an assurance case is only one component in a broader
commitment to ethical reflection, deliberation and participatory design,
development, and deployment. Such processes can only flourish in
organisations and teams that take care to build a culture of readiness,
reflection, and responsible research and innovation. This was evidently
not the case with the production of the Nimrod safety case, as the
review goes on to acknowledge that the above matters raised ``question
marks about the prevailing \emph{ethical culture} at BAE Systems''
(\protect\hyperlink{ref-haddon-cave2009}{Haddon-Cave et al. 2009, 10}).
Others have also recognised the importance of organisational
culture---construed broadly as both interpersonal norms and values
(e.g.~culture of reflection), as well as external constraints or
practical mechanisms (e.g.~incident reporting systems). For example, in
discussing the application of safety cases to healthcare, Sujan and
Habli (\protect\hyperlink{ref-sujan2021}{2021, 4}) are optimistic about
the possibility of using safety cases for improving digital health
innovations, but only when accompanied by ``far-reaching structural
changes.''

In addition to recognisiong the importance of organisational culture,
Sujan and Habli (\protect\hyperlink{ref-sujan2021}{2021}) also offer two
reasons why there is no conclusive evidence for the \emph{efficacy} of
safety assurance more generally. First, safety cases are typically
employed as regulatory instruments in ``high-hazard settings,''
characterised by high-severity but low-frequency events (e.g.~the
catastrophic failure and loss of an aircraft, as above). As such, it is
difficult to obtain statistically meaningful data about the casual
impact that safety cases have. Second, the process of developing a
safety case varies widely across domains (e.g.~nuclear, aviation),
making it hard to determine which factors of a safety case or the
supporting culture are causally relevant to increasing or decreasing
safety risks.

While these challenges have been raised in connection with safety cases
specifically, it is reasonable to conclude that similar challenges will
apply to the possible adoption and uptake of ethical assurance cases.
For example, in their critique of the ideal of transparency as it
applies to algorithmic systems, Annany and Crawford
(\protect\hyperlink{ref-ananny2018}{2018, 980}) question the assumption
that transparency engenders trust in organisations and systems, given
the lack of confirmatory empirical research. As a next step, it will be
necessary to identify and determine how different substantive and
structural factors may contribute to the success or failure of ethical
assurance.

\hypertarget{systems-and-standards}{%
\subsubsection{Systems and Standards}\label{systems-and-standards}}

In addition to the directions implied by the previous two open
questions, there is a further avenue that would help develop the current
proposals into a more robust research programme: practical systems and
standards that can help teams and organisations implement ethical
assurance.

Ethical assurance cases could end up being large and complex, especially
when mutliple goals are interlinked (see
\protect\hyperlink{top-level-normative-goal}{§4.3.1}). Therefore, it
would be sensible to consider---as happens in ABA more generally---how
software tools can support the process of both building and interacting
with an ethical assurance case. For example, could online software
platforms help teams iteratively construct an ethical assurance case and
also open it up to different groups of stakeholders?

To avoid the situation where we end up with a miscellany of tools and
approaches, much like the problem we described at the start of this
article, it would be wise to consider whether there are pre-existing
standards and best practices to draw from (e.g.~interoperability
standards). For ABA more generally, the Object Management Group
(OMG)---a computer industry standards consortium---has developed the
Structured Assurance Case Metamodel (SACM), which is used to represent
assurance case (\protect\hyperlink{ref-objectmanagementgroup2018}{Object
Management Group Macrh 2018}). It also provides further constraints on
the use of, say, argument patterns and relevant evidence, or acceptable
syntax and semantics. As Hawkins et al.
(\protect\hyperlink{ref-hawkins2015}{2015}) acknowledge, having a
(meta)model of the assurance process can bring certain benefits of
model-driven approaches to engineering, such as automation, traceability
and accountability, transformation and validation. The value of this
approach, however, is conditional on the extent to which it is adopted
and supported by the community (e.g.~used by developers, recognised by
regulators). While we have avoided the issue of whether ethical
assurance could conform to such a standard, there appears to be,
\emph{prima facie}, no reason why a model-based approach could not be
pursued, in order to build, for instance, a metamodel that synthesises
the reflect, act, and justify model (see Figure 3) with the project
lifecycle model (see Figure 2) and informal logic of an ethical
assurance case (see Figure 8).

Pursuing such an approach could also increase the value of possible
online software tools. For example, consider a metamodel that provided
guidance on the adoption of argument patterns for ethical assurance
cases that made use of multiple top-level normative goals. If particular
goals (and supporting claims and evidence) were of more interest to a
specific group of stakeholders (e.g.~auditors), then the metamodel could
enable the use of simple software filters that could be applied when
interacting with an ethical assurance case, allowing the stakeholder
group to focus in on only those parts of the argument that were
relevant. Or, it could support the reframing of an argument at a
different level of complexity when certain evidential claims
(e.g.~details about an ML component) require technical expertise that go
beyond the scope of a certain group of stakeholders.

These capabilities would also depend on the development of ethical
argument patterns. However, by developing a series of reusable templates
(i.e.~patterns) that have proved to be helpful in supporting criticial
reflection and deliberation in specific contexts or use cases, the
development of ethical assurance cases could be made more efficient and
effective. For instance, it would be possible to develop software tools
linked to a structured repository of argument patterns (e.g.~through an
API), which prompt users with a series of questions about which concept
of `fairness' they are using, and offer guides for specifying high-level
ethical principles in a specific context.

Therefore, an important next step here is to consider the formal and
syntactical representation of ethical assurance in more detail. For
present purposes, we have sidestepped issues such as which notation
schema it will rely upon (e.g.~GSN) and whether it will conform to
existing standards (e.g.~SACM) in order to focus on the broader purpose
and motivation of the ethical assurance methodology. This gap will, of
course, need to be addressed to ensure the full potential of ethical
assurance is realised. For the time being, however, it is sufficient to
conclude by noting that there are good reasons to believe that the
development of systems and standards is a worthwhile avenue to explore.
We hope that by outlinining both the potential value of ethical
assurance in general, and outlining concrete opportunities for its
development, that an active community will emerge to help realise its
potential for ensuring that data-science and AI contribute to an
inclusive and collective social good.

\textbf{Conflict of Interest Statement}

On behalf of all authors, the corresponding author states that there is
no conflict of interest.

\textbf{Funding \& Acknowledgements}

This research was supported by a grant from the UKRI Trustworthy
Autonomous Systems Hub, awarded to Dr Christopher Burr.

We wish to thank Ibrahim Habli, Mike Katell, and Geoff Keeling for their
insightful comments on earlier drafts of this article, as well as
offering suggestions for further research that took the article in
valuable directions, which it otherwise would not have explored.

\hypertarget{references}{%
\section*{References}\label{references}}
\addcontentsline{toc}{section}{References}

\hypertarget{refs}{}
\begin{CSLReferences}{1}{0}
\leavevmode\hypertarget{ref-ananny2018}{}%
Ananny, Mike, and Kate Crawford. 2018. {``Seeing Without Knowing:
Limitations of the Transparency Ideal and Its Application to Algorithmic
Accountability.''} \emph{New Media \& Society} 20 (3): 973--89.
\url{https://doi.org/10.1177/1461444816676645}.

\leavevmode\hypertarget{ref-andersson2013}{}%
Andersson, Edward, Sam McLean, Metin Parlak, and Gabrielle Melvin. 2013.
{``From {Fairy Tale} to {Reality}{{Dispelling}} the {Myths Around
Citizen Engagement}.''} {Involve and the RSA}.

\leavevmode\hypertarget{ref-arnold2019}{}%
Arnold, M., R. K. E. Bellamy, M. Hind, S. Houde, S. Mehta, A.
Mojsilović, R. Nair, et al. 2019. {``{FactSheets}: Increasing Trust in
{AI} Services Through Supplier's Declarations of Conformity.''}
\emph{IBM Journal of Research and Development} 63 (4/5): 6:1--13.
\url{https://doi.org/10.1147/JRD.2019.2942288}.

\leavevmode\hypertarget{ref-ashmore2019}{}%
Ashmore, Rob, Radu Calinescu, and Colin Paterson. 2019. {``Assuring the
{Machine Learning Lifecycle}: Desiderata, {Methods}, and
{Challenges}.''} \emph{arXiv:1905.04223 {[}Cs, Stat{]}}, May.
\url{http://arxiv.org/abs/1905.04223}.

\leavevmode\hypertarget{ref-beauchamp2004}{}%
Beauchamp, Tom L., and David DeGrazia. 2004. {``Principles and
{Principlism}.''} In \emph{Handbook of {Bioethics}}, edited by George
Khushf, 55--74. {Dordrecht}: {Springer Netherlands}.
\url{https://doi.org/10.1007/1-4020-2127-5_3}.

\leavevmode\hypertarget{ref-beauchamp2013}{}%
Beauchamp, Tom L, and James F Childress. 2013. \emph{Principles of
Biomedical Ethics}. Seventh. {New York, N.Y.}: {Oxford University
Press}.

\leavevmode\hypertarget{ref-bender2018}{}%
Bender, Emily M., and Batya Friedman. 2018. {``Data {Statements} for
{Natural Language Processing}: Toward {Mitigating System Bias} and
{Enabling Better Science}.''} \emph{Transactions of the Association for
Computational Linguistics} 6: 587--604.
\url{https://doi.org/10.1162/tacl_a_00041}.

\leavevmode\hypertarget{ref-benjamin2019}{}%
Benjamin, Ruha. 2019. \emph{Race After Technology: Abolitionist Tools
for the New {Jim} Code}. {Medford, MA}: {Polity}.

\leavevmode\hypertarget{ref-binns2018}{}%
Binns, Reuben. 2018. {``What {Can Political Philosophy Teach Us} about
{Algorithmic Fairness}?''} \emph{IEEE Secur. Privacy} 16 (3): 73--80.
\url{https://doi.org/10.1109/MSP.2018.2701147}.

\leavevmode\hypertarget{ref-bloomfield2010}{}%
Bloomfield, Robin, and Peter Bishop. 2010. {``Safety and {Assurance
Cases}: Past, {Present} and {Possible Future} {} an {Adelard
Perspective}.''} In \emph{Making {Systems Safer}}, edited by Chris Dale
and Tom Anderson, 51--67. {London}: {Springer London}.
\url{https://doi.org/10.1007/978-1-84996-086-1_4}.

\leavevmode\hypertarget{ref-brundage2020}{}%
Brundage, Miles, Shahar Avin, Jasmine Wang, Haydn Belfield, Gretchen
Krueger, Gillian Hadfield, Heidy Khlaaf, et al. 2020. {``Toward
{Trustworthy AI Development}: Mechanisms for {Supporting Verifiable
Claims}.''} \emph{arXiv:2004.07213 {[}Cs{]}}, April.
\url{http://arxiv.org/abs/2004.07213}.

\leavevmode\hypertarget{ref-burton2020}{}%
Burton, Simon, Ibrahim Habli, Tom Lawton, John McDermid, Phillip Morgan,
and Zoe Porter. 2020. {``Mind the Gaps: Assuring the Safety of
Autonomous Systems from an Engineering, Ethical, and Legal
Perspective.''} \emph{Artificial Intelligence} 279 (February): 103201.
\url{https://doi.org/10.1016/j.artint.2019.103201}.

\leavevmode\hypertarget{ref-calinescu2018}{}%
Calinescu, Radu, Danny Weyns, Simos Gerasimou, Muhammad Usman Iftikhar,
Ibrahim Habli, and Tim Kelly. 2018. {``Engineering {Trustworthy
Self}-{Adaptive Software} with {Dynamic Assurance Cases}.''} \emph{IEEE
Transactions on Software Engineering} 44 (11): 1039--69.
\url{https://doi.org/10.1109/TSE.2017.2738640}.

\leavevmode\hypertarget{ref-cartwright2012}{}%
Cartwright, Nancy, and Jeremy Hardie. 2012. \emph{Evidence-Based Policy:
A Practical Guide to Doing It Better}. {Oxford University Press}.

\leavevmode\hypertarget{ref-cleland2012}{}%
Cleland, George M, Ibrahim Habli, John Medhurst, and Health Foundation
(Great Britain). 2012. {``Evidence: Using Safety Cases in Industry and
Healthcare.''}

\leavevmode\hypertarget{ref-cobbe2021}{}%
Cobbe, Jennifer, Michelle Seng Ah Lee, and Jatinder Singh. 2021.
{``Reviewable {Automated Decision}-{Making}: A {Framework} for
{Accountable Algorithmic Systems}.''} In \emph{Proceedings of the 2021
{ACM Conference} on {Fairness}, {Accountability}, and {Transparency}},
598--609. {Virtual Event Canada}: {ACM}.
\url{https://doi.org/10.1145/3442188.3445921}.

\leavevmode\hypertarget{ref-collingridge1980}{}%
Collingridge, David. 1980. \emph{The Social Control of Technology}. {New
York}: {St. Martin's Press}.

\leavevmode\hypertarget{ref-collins2015a}{}%
Collins, Gary S., Johannes B. Reitsma, Douglas G. Altman, and Karel G.
M. Moons. 2015. {``Transparent {Reporting} of a Multivariable Prediction
Model for {Individual Prognosis Or Diagnosis} ({TRIPOD}): The {TRIPOD
Statement}.''} \emph{Ann Intern Med} 162 (1): 55.
\url{https://doi.org/10.7326/M14-0697}.

\leavevmode\hypertarget{ref-lawcommission2020}{}%
Commission, Law. 2020. {``Automated {Vehicles}: Summary of {Consultation
Paper} 3 {} {A} Regulatory Framework for Automated Vehicles.''}

\leavevmode\hypertarget{ref-gsncommunity2018}{}%
Community, GSN. 2018. {``{GSN Community Standard} ({Version} 2).''} {The
Assurance Case Working Group}.

\leavevmode\hypertarget{ref-diakopoulos2014}{}%
Diakopoulos, Nicholas. 2014. {``Algorithmic {Accountability Reporting}:
On the {Investigation} of {Black Boxes}.''} {Tow Center for Digital
Journalism}.

\leavevmode\hypertarget{ref-diakopoulos2015}{}%
---------. 2015. {``Algorithmic {Accountability}: Journalistic
Investigation of Computational Power Structures.''} \emph{Digital
Journalism} 3 (3): 398--415.
\url{https://doi.org/10.1080/21670811.2014.976411}.

\leavevmode\hypertarget{ref-dryzek2003}{}%
Dryzek, John S., and Christian List. 2003. {``Social {Choice Theory} and
{Deliberative Democracy}: A {Reconciliation}.''} \emph{British Journal
of Political Science} 33 (1): 1--28.

\leavevmode\hypertarget{ref-eemeren2004}{}%
Eemeren, Frans H Van, and Rob Grootendorst. 2004. \emph{A {Systematic
Theory} of {Argumentation}: The Pragma-Dialectical Approach}. {Cambridge
University Press}.

\leavevmode\hypertarget{ref-fang2020}{}%
Fang, Huanming, and Hui Miao. 2020. {``Introducing the {Model Card
Toolkit} for {Easier Model Transparency Reporting}.''} \emph{Google AI
Blog}.

\leavevmode\hypertarget{ref-gebru2018}{}%
Gebru, Timnit, Jamie Morgenstern, Briana Vecchione, Jennifer Wortman
Vaughan, Hanna Wallach, Hal Daumé III, and Kate Crawford. 2018.
{``Datasheets for {Datasets}.''} In \emph{Proceedings of the 5th
{Workshop} on {Fairness}, {Accountability}, and {Transparency} in
{Machine Learning}}. \url{http://arxiv.org/abs/1803.09010}.

\leavevmode\hypertarget{ref-gebru2019}{}%
---------. 2019. {``Datasheets for {Datasets}.''} \emph{arXiv:1803.09010
{[}Cs{]}}. \url{http://arxiv.org/abs/1803.09010}.

\leavevmode\hypertarget{ref-habermas1998}{}%
Habermas, Jürgen. 1998. \emph{On the Pragmatics of Communication}. {MIT
press}.

\leavevmode\hypertarget{ref-habli2021}{}%
Habli, Ibrahim, Rob Alexander, and Richard Hawkins. 2021. {``Safety
{Cases}: An {Impending Crisis}?''} In \emph{Safety-{Critical Systems
Symposium} ({SSS}'21)}, 18.

\leavevmode\hypertarget{ref-habli2020}{}%
Habli, Ibrahim, Rob Alexander, Richard Hawkins, Mark Sujan, John
McDermid, Chiara Picardi, and Tom Lawton. 2020. {``Enhancing {COVID}-19
Decision Making by Creating an Assurance Case for Epidemiological
Models.''} \emph{BMJ Health \&Amp; Care Informatics} 27 (3): e100165.
\url{https://doi.org/10.1136/bmjhci-2020-100165}.

\leavevmode\hypertarget{ref-haddon-cave2009}{}%
Haddon-Cave, Charles, Great Britain, Parliament, and House of Commons.
2009. {``{THE NIMROD REVIEW}: An Independent Review into the Broader
Issues Surrounding the Loss of the {RAF Nimrod Mr2 Aircraft Xv230} in
{Afghanistan} in 2006.''} {London}: {Stationery Office}.

\leavevmode\hypertarget{ref-hao2019}{}%
Hao, Karen. 2019. {``In 2020, Let's Stop {AI} Ethics-Washing and
Actually Do Something.''} \emph{MIT Technology Review}.
https://www.technologyreview.com/2019/12/27/57/ai-ethics-washing-time-to-act/.

\leavevmode\hypertarget{ref-hawkins2009}{}%
Hawkins, R. D., and T. P. Kelly. 2009. {``Software Safety Assurance -
What Is Sufficient?''} In \emph{4th {IET International Conference} on
{Systems Safety} 2009. {Incorporating} the {SaRS Annual Conference}},
2A3--3. {London, UK}: {IET}. \url{https://doi.org/10.1049/cp.2009.1542}.

\leavevmode\hypertarget{ref-hawkins2015}{}%
Hawkins, Richard, Ibrahim Habli, Dimitris Kolovos, Richard Paige, and
Tim Kelly. 2015. {``Weaving an {Assurance Case} from {Design}: A
{Model}-{Based Approach}.''} In \emph{2015 {IEEE} 16th {International
Symposium} on {High Assurance Systems Engineering}}, 110--17. {Daytona
Beach Shores, FL, USA}: {IEEE}.
\url{https://doi.org/10.1109/HASE.2015.25}.

\leavevmode\hypertarget{ref-hawkins2011}{}%
Hawkins, Richard, Tim Kelly, John Knight, and Patrick Graydon. 2011.
{``A {New Approach} to Creating {Clear Safety Arguments}.''} In
\emph{Advances in {Systems Safety}}, edited by Chris Dale and Tom
Anderson, 3--23. {London}: {Springer London}.
\url{https://doi.org/10.1007/978-0-85729-133-2_1}.

\leavevmode\hypertarget{ref-hawkins2021}{}%
Hawkins, Richard, Colin Paterson, Chiara Picardi, Yan Jia, Radu
Calinescu, and Ibrahim Habli. 2021. {``Guidance on the {Assurance} of
{Machine Learning} in {Autonomous Systems}.''} {University of York}:
{Assuring Autonomy International Programme (AAIP)}.

\leavevmode\hypertarget{ref-high-levelexpertgrouponai2019}{}%
High-Level Expert Group on AI. 2019. {``Ethics {Guidelines} for
{Trustworthy AI}.''} {European Commission}.

\leavevmode\hypertarget{ref-ho2015}{}%
Ho, Hock Lai. 2015. {``The {Legal Concept} of {Evidence}.''} In
\emph{The {Stanford Encyclopedia} of {Philosophy}}, edited by Edward N.
Zalta, Winter 2015. {Metaphysics Research Lab, Stanford University}.

\leavevmode\hypertarget{ref-holland2018}{}%
Holland, Sarah, Ahmed Hosny, Sarah Newman, Joshua Joseph, and Kasia
Chmielinski. 2018. \emph{The {Dataset Nutrition Label}: A {Framework To
Drive Higher Data Quality Standards}}.

\leavevmode\hypertarget{ref-horty2014}{}%
Horty, John F. 2014. \emph{Reasons as Defaults}. {New-York (N.Y.)}:
{Oxford University Press}.

\leavevmode\hypertarget{ref-ico2020b}{}%
ICO. 2020. {``Guidance on the {AI} Auditing Framework.''} {Information
Commissioner's Office}.

\leavevmode\hypertarget{ref-ico2020}{}%
ICO, and Alan Turing Institute. 2020. {``Explaining Decisions Made with
{AI}.''}

\leavevmode\hypertarget{ref-kalluri2020}{}%
Kalluri, Pratyusha. 2020. {``Don't Ask If Artificial Intelligence Is
Good or Fair, Ask How It Shifts Power.''} \emph{Nature} 583 (7815):
169--69. \url{https://doi.org/10.1038/d41586-020-02003-2}.

\leavevmode\hypertarget{ref-kelly1998}{}%
Kelly, Timothy Patrick. 1998. {``Arguing {Safety} {} {A Systematic
Approach} to {Managing Safety Cases}.''} PhD thesis, {Department of
Computer Science}: University of York.

\leavevmode\hypertarget{ref-kind2020}{}%
Kind, Carly. 2020. {``The Term {`Ethical {AI}'} Is Finally Starting to
Mean Something \textbar{} {VentureBeat}.''} \emph{VentureBeat}.
https://venturebeat.com/2020/08/23/the-term-ethical-ai-is-finally-starting-to-mean-something/.

\leavevmode\hypertarget{ref-kroll2021}{}%
Kroll, Joshua A. 2021. {``Outlining {Traceability}: A {Principle} for
{Operationalizing Accountability} in {Computing Systems}.''} In
\emph{Proceedings of the 2021 {ACM Conference} on {Fairness},
{Accountability}, and {Transparency}}, 758--71. {Virtual Event Canada}:
{ACM}. \url{https://doi.org/10.1145/3442188.3445937}.

\leavevmode\hypertarget{ref-leslie2019}{}%
Leslie, David. 2019. {``Understanding Artificial Intelligence Ethics and
Safety.''} {The Alan Turing Institute}.

\leavevmode\hypertarget{ref-leslie2020a}{}%
---------. 2020. {``The Secret Life of Algorithms in the Time of
{COVID}-19.''} \emph{The Alan Turing Institute}.
https://www.turing.ac.uk/blog/secret-life-algorithms-time-covid-19.

\leavevmode\hypertarget{ref-leslie2021}{}%
---------. 2021. {``The {Arc} of the {Data Scientific Universe}.''}
\emph{Harvard Data Science Review}, January.
\url{https://doi.org/10.1162/99608f92.938a18d7}.

\leavevmode\hypertarget{ref-lucyk2017}{}%
Lucyk, Kelsey, and Lindsay McLaren. 2017. {``Taking Stock of the Social
Determinants of Health: A Scoping Review.''} Edited by Spencer Moore.
\emph{PLoS ONE} 12 (5): e0177306.
\url{https://doi.org/10.1371/journal.pone.0177306}.

\leavevmode\hypertarget{ref-lundberg2020}{}%
Lundberg, Scott. 2020. {``Slundberg/Shap.''}

\leavevmode\hypertarget{ref-maksimov2018}{}%
Maksimov, Mike, Nick L. S. Fung, Sahar Kokaly, and Marsha Chechik. 2018.
{``Two {Decades} of {Assurance Case Tools}: A {Survey}.''} In
\emph{Developments in {Language Theory}}, edited by Mizuho Hoshi and
Shinnosuke Seki, 11088:49--59. {Cham}: {Springer International
Publishing}. \url{https://doi.org/10.1007/978-3-319-99229-7_6}.

\leavevmode\hypertarget{ref-mitchell2019}{}%
Mitchell, Margaret, Simone Wu, Andrew Zaldivar, Parker Barnes, Lucy
Vasserman, Ben Hutchinson, Elena Spitzer, Inioluwa Deborah Raji, and
Timnit Gebru. 2019. {``Model {Cards} for {Model Reporting}.''}
\emph{Proceedings of the Conference on Fairness, Accountability, and
Transparency - FAT* '19}, 220--29.
\url{https://doi.org/10.1145/3287560.3287596}.

\leavevmode\hypertarget{ref-moretti2013}{}%
Moretti, Luca, and Tommaso Piazza. 2013. {``Transmission of
{Justification} and {Warrant},''} November.

\leavevmode\hypertarget{ref-morley2019f}{}%
Morley, Jessica, Luciano Floridi, Libby Kinsey, and Anat Elhalal. 2019.
{``From {What} to {How}: An {Initial Review} of {Publicly Available AI
Ethics Tools}, {Methods} and {Research} to {Translate Principles} into
{Practices}.''} \emph{Sci Eng Ethics}, December.
\url{https://doi.org/10.1007/s11948-019-00165-5}.

\leavevmode\hypertarget{ref-mokander2021}{}%
Mökander, Jakob, and Luciano Floridi. 2021. {``Ethics-{Based Auditing}
to {Develop Trustworthy AI}.''} \emph{Minds \& Machines}, February.
\url{https://doi.org/10.1007/s11023-021-09557-8}.

\leavevmode\hypertarget{ref-oneill2002}{}%
O'Neill, Onora. 2002. \emph{A {Question} of {Trust}}. {Cambridge}:
{Cambridge University Press}.

\leavevmode\hypertarget{ref-objectmanagementgroup2018}{}%
Object Management Group, Adelard. Macrh 2018. {``Structured {Assurance
Case Metamodel} ({SACM}) {Version} 2.0.''}

\leavevmode\hypertarget{ref-owen2013}{}%
Owen, Richard, J. R. Bessant, and Maggy Heintz, eds. 2013.
\emph{Responsible Innovation}. {Chichester, West Sussex, United
Kingdom}: {Wiley}.

\leavevmode\hypertarget{ref-pair2020}{}%
PAIR. 2020. {``What-{If Tool} - {People} + {AI Research} ({PAIR}).''}
https://pair-code.github.io/what-if-tool/.

\leavevmode\hypertarget{ref-picardi2020}{}%
Picardi, Chiara, Colin Paterson, Richard Hawkins, Radu Calinescu, and
Ibrahim Habli. 2020. {``Assurance {Argument Patterns} and {Processes}
for {Machine Learning} in {Safety}-{Related Systems}.''} In
\emph{Proceedings of the {Workshop} on {Artificial Intelligence Safety}
({SafeAI} 2020)}, 23--30. {CEUR Workshop Proceedings}. {CEUR Workshop
Proceedings}.

\leavevmode\hypertarget{ref-raji2020}{}%
Raji, Inioluwa Deborah, Andrew Smart, Rebecca N White, Margaret
Mitchell, Timnit Gebru, Ben Hutchinson, Jamila Smith-Loud, Daniel
Theron, and Parker Barnes. 2020. {``Closing the {AI Accountability Gap}:
Defining an {End}-to-{End Framework} for {Internal Algorithmic
Auditing},''} 12.

\leavevmode\hypertarget{ref-rawls1999}{}%
Rawls, John. 1999. \emph{A {Theory} of {Justice}}. Revised Edition.
{Cambridge, Mass}: {Belknap Press of Harvard University Press}.

\leavevmode\hypertarget{ref-reisman2018}{}%
Reisman, Dillon, Jason Schultz, Kate Crawford, and Meredith Whittaker.
2018. {``Algorithmic {Impact Assessments}: A {Practical Framework} for
{Public Accountability}.''} {AI Now}.

\leavevmode\hypertarget{ref-ibmresearch2018}{}%
Research, IBM. 2018. {``Introducing {AI Fairness} 360, {A Step Towards
Trusted AI}.''} \emph{IBM Research Blog}.
https://www.ibm.com/blogs/research/2018/09/ai-fairness-360/.

\leavevmode\hypertarget{ref-royalcollegeofphysicians2017}{}%
Royal College of Physicians. 2017. {``National {Early Warning Score}
({NEWS}) 2.''} \emph{RCP London}.
https://www.rcplondon.ac.uk/projects/outputs/national-early-warning-score-news-2.

\leavevmode\hypertarget{ref-selbst2019}{}%
Selbst, Andrew D., Danah Boyd, Sorelle A. Friedler, Suresh
Venkatasubramanian, and Janet Vertesi. 2019. {``Fairness and
{Abstraction} in {Sociotechnical Systems}.''} In \emph{Proceedings of
the {Conference} on {Fairness}, {Accountability}, and {Transparency} -
{FAT}* '19}, 59--68. {Atlanta, GA, USA}: {ACM Press}.
\url{https://doi.org/10.1145/3287560.3287598}.

\leavevmode\hypertarget{ref-stilgoe2013}{}%
Stilgoe, Jack, Richard Owen, and Phil Macnaghten. 2013. {``Developing a
Framework for Responsible Innovation.''} \emph{Research Policy} 42 (9):
1568--80. \url{https://doi.org/10.1016/j.respol.2013.05.008}.

\leavevmode\hypertarget{ref-sujan2021}{}%
Sujan, Mark, and Ibrahim Habli. 2021. {``Safety Cases for Digital Health
Innovations: Can They Work?''} \emph{BMJ Qual Saf}, May,
bmjqs-2021-012983. \url{https://doi.org/10.1136/bmjqs-2021-012983}.

\leavevmode\hypertarget{ref-sweenor2020}{}%
Sweenor, David, Steven Hillion, Dan Rope, Dev Kannabiran, Thomas Hill,
Michael O'Connell, and an O'Reilly Media Company Safari. 2020. \emph{{ML
Ops}: Operationalizing {Data Science}}.

\leavevmode\hypertarget{ref-toulmin2003}{}%
Toulmin, Stephen. 2003. \emph{The {Uses} of {Argument}, {Updated
Edition}}. {Cambridge}: {Cambridge University Press}.

\leavevmode\hypertarget{ref-ward2020}{}%
Ward, Francis Rhys, and Ibrahim Habli. 2020. {``An {Assurance Case
Pattern} for the {Interpretability} of {Machine Learning} in
{Safety}-{Critical Systems}.''} In \emph{Computer {Safety},
{Reliability}, and {Security}. {SAFECOMP} 2020 {Workshops}}, edited by
António Casimiro, Frank Ortmeier, Erwin Schoitsch, Friedemann Bitsch,
and Pedro Ferreira, 12235:395--407. {Cham}: {Springer International
Publishing}. \url{https://doi.org/10.1007/978-3-030-55583-2_30}.

\end{CSLReferences}

\end{document}